\documentclass[twocolumn]{aa}

\usepackage{graphicx}
\usepackage{amsmath,amsfonts,amssymb}
\usepackage{txfonts}
\usepackage{color}
\usepackage{natbib}
\usepackage{float}
\usepackage{dblfloatfix}
\usepackage{afterpage}
\usepackage{ifthen}
\usepackage[morefloats=12]{morefloats}
\usepackage{placeins}
\usepackage{multicol}
\bibpunct{(}{)}{;}{a}{}{,}
\usepackage[switch]{lineno}
\definecolor{linkcolor}{rgb}{0.6,0,0}
\definecolor{citecolor}{rgb}{0,0,0.75}
\definecolor{urlcolor}{rgb}{0.12,0.46,0.7}
\usepackage[breaklinks, colorlinks, urlcolor=urlcolor,
    linkcolor=linkcolor,citecolor=citecolor,pdfencoding=auto]{hyperref}
\hypersetup{linktocpage}
\usepackage{bold-extra}
\usepackage{xcolor}

\def\setsymbol#1#2{\expandafter\def\csname #1\endcsname{#2}}
\def\getsymbol#1{\csname #1\endcsname}

\def\Planck{\textit{Planck}}

\newbox\tablebox    \newdimen\tablewidth
\def\leaderfil{\leaders\hbox to 5pt{\hss.\hss}\hfil}
\def\tablenote#1 #2\par{\begingroup \parindent=0.8em
    \abovedisplayshortskip=0pt\belowdisplayshortskip=0pt
    \noindent
    $$\hss\vbox{\hsize\tablewidth \hangindent=\parindent \hangafter=1 \noindent
    \hbox to \parindent{$^#1$\hss}\strut#2\strut\par}\hss$$
    \endgroup}

\def\L2{\ifmmode L_2\else $L_2$\fi}

\def\DeltaT{\ifmmode \Delta T\else $\Delta T$\fi}
\def\deltat{\ifmmode \Delta t\else $\Delta t$\fi}
\def\fknee{\ifmmode f_{\rm knee}\else $f_{\rm knee}$\fi}
\def\Fmax{\ifmmode F_{\rm max}\else $F_{\rm max}$\fi}
\def\solar{\ifmmode{\rm M}_{\mathord\odot}\else${\rm M}_{\mathord\odot}$\fi}
\def\Msolar{\ifmmode{\rm M}_{\mathord\odot}\else${\rm M}_{\mathord\odot}$\fi}
\def\Lsolar{\ifmmode{\rm L}_{\mathord\odot}\else${\rm L}_{\mathord\odot}$\fi}
\def\inv{\ifmmode^{-1}\else$^{-1}$\fi}
\def\mo{\ifmmode^{-1}\else$^{-1}$\fi}
\def\sup#1{\ifmmode ^{\rm #1}\else $^{\rm #1}$\fi}
\def\expo#1{\ifmmode \times 10^{#1}\else $\times 10^{#1}$\fi}
\def\,{\thinspace}
\def\lsim{\mathrel{\raise .4ex\hbox{\rlap{$<$}\lower 1.2ex\hbox{$\sim$}}}}
\def\gsim{\mathrel{\raise .4ex\hbox{\rlap{$>$}\lower 1.2ex\hbox{$\sim$}}}}

\def\simprop{\mathrel{\raise .4ex\hbox{\rlap{$\propto$}\lower 1.2ex\hbox{$\sim$}}}}
\def\deg{\ifmmode^\circ\else$^\circ$\fi}
\def\pdeg{\ifmmode $\setbox0=\hbox{$^{\circ}$}\rlap{\hskip.11\wd0 .}$^{\circ}
          \else \setbox0=\hbox{$^{\circ}$}\rlap{\hskip.11\wd0 .}$^{\circ}$\fi}
\def\arcs{\ifmmode {^{\scriptstyle\prime\prime}}
          \else $^{\scriptstyle\prime\prime}$\fi}
\def\arcm{\ifmmode {^{\scriptstyle\prime}}
          \else $^{\scriptstyle\prime}$\fi}
\newdimen\sa  \newdimen\sb
\def\parcs{\sa=.07em \sb=.03em
     \ifmmode \hbox{\rlap{.}}^{\scriptstyle\prime\kern -\sb\prime}\hbox{\kern -\sa}
     \else \rlap{.}$^{\scriptstyle\prime\kern -\sb\prime}$\kern -\sa\fi}
\def\parcm{\sa=.08em \sb=.03em
     \ifmmode \hbox{\rlap{.}\kern\sa}^{\scriptstyle\prime}\hbox{\kern-\sb}
     \else \rlap{.}\kern\sa$^{\scriptstyle\prime}$\kern-\sb\fi}
\def\ra[#1 #2 #3.#4]{#1\sup{h}#2\sup{m}#3\sup{s}\llap.#4}
\def\dec[#1 #2 #3.#4]{#1\deg#2\arcm#3\arcs\llap.#4}
\def\deco[#1 #2 #3]{#1\deg#2\arcm#3\arcs}
\def\rra[#1 #2]{#1\sup{h}#2\sup{m}}

\def\dots{\relax\ifmmode \ldots\else $\ldots$\fi}
\def\WHzsr{\ifmmode $W\,Hz\mo\,sr\mo$\else W\,Hz\mo\,sr\mo\fi}
\def\mHz{\ifmmode $\,mHz$\else \,mHz\fi}
\def\GHz{\ifmmode $\,GHz$\else \,GHz\fi}
\def\mKs{\ifmmode $\,mK\,s$^{1/2}\else \,mK\,s$^{1/2}$\fi}
\def\muKs{\ifmmode \,\mu$K\,s$^{1/2}\else \,$\mu$K\,s$^{1/2}$\fi}
\def\muKRJs{\ifmmode \,\mu$K$_{\rm RJ}$\,s$^{1/2}\else \,$\mu$K$_{\rm RJ}$\,s$^{1/2}$\fi}
\def\muKHz{\ifmmode \,\mu$K\,Hz$^{-1/2}\else \,$\mu$K\,Hz$^{-1/2}$\fi}
\def\MJysr{\ifmmode \,$MJy\,sr\mo$\else \,MJy\,sr\mo\fi}
\def\MJysrmK{\ifmmode \,$MJy\,sr\mo$\,mK$_{\rm CMB}\mo\else \,MJy\,sr\mo\,mK$_{\rm CMB}\mo$\fi}
\def\microns{\ifmmode \,\mu$m$\else \,$\mu$m\fi}

\def\muK{\ifmmode \,\mu$K$\else \,$\mu$\hbox{K}\fi}
\def\microK{\ifmmode \,\mu$K$\else \,$\mu$\hbox{K}\fi}
\def\muW{\ifmmode \,\mu$W$\else \,$\mu$\hbox{W}\fi}
\def\kms{\ifmmode $\,km\,s$^{-1}\else \,km\,s$^{-1}$\fi}
\def\kmsMpc{\ifmmode $\,\kms\,Mpc\mo$\else \,\kms\,Mpc\mo\fi}
\providecommand{\sorthelp}[1]{}

\def\WMAP{\emph{WMAP}}
\def\COBE{\emph{COBE}}
\def\wmap{\emph{WMAP}}

\def\commander{\texttt{Commander}}

\renewcommand{\d}[0]{\vec{d}}

\newcommand{\A}[0]{\mathrm{A}}
\newcommand{\B}[0]{\mathrm{B}}

\newcommand{\n}[0]{\vec{n}}

\newcommand{\s}[0]{\vec{s}}
\renewcommand{\a}[0]{\vec{a}}

\newcommand{\T}[0]{\tens{T}}

\renewcommand{\L}[0]{\tens{L}}
\newcommand{\g}[0]{\vec{g}}
\newcommand{\N}[0]{\tens{N}}

\renewcommand{\r}[0]{\vec{r}}

\renewcommand{\P}[0]{\tens{P}}

\newcommand{\BP}{\textsc{BeyondPlanck}}
\newcommand{\cosmoglobe}{\textsc{Cosmoglobe}}

\newcommand{\K}[0]{\textit K}
\newcommand{\Ka}[0]{\textit{Ka}}
\newcommand{\Q}[0]{\textit Q}
\newcommand{\V}[0]{\textit V}
\newcommand{\W}[0]{\textit W}
\newcommand{\e}{\mathrm e}
\newcommand{\cvar}{\ensuremath{c(\vartheta, \varphi, \psi)}}

\newcommand{\ncorr}{\vec n_\mathrm{corr}}
\newcommand{\Dbp}{\Delta_\mathrm{bp}}

\def\inv{^{-1}}

\begin{document}

\title{\bfseries{From \scshape{BeyondPlanck} to \scshape{Cosmoglobe}}: \\Preliminary \textit{WMAP} $Q$-band analysis}
%This author list corresponds to \title{Author list for L04\_CMB\_Foregrounds\_Extraction}
%Prepared by M. Lopez-Caniego (Marcos.Lopez.Caniego@sciops.esa.int), ESAC/ESA
%This version is from Thu Jul 12 18:11:48 2018 CET
%\subtitle{There are 152 co-authors in this list}
\newcommand{\oslo}[0]{1}
\newcommand{\milanoA}[0]{2}
\newcommand{\milanoB}[0]{3}
\newcommand{\milanoC}[0]{4}
\newcommand{\triesteB}[0]{5}
\newcommand{\planetek}[0]{6}
\newcommand{\princeton}[0]{7}
\newcommand{\jpl}[0]{8}
\newcommand{\helsinkiA}[0]{9}
\newcommand{\helsinkiB}[0]{10}
\newcommand{\nersc}[0]{11}
\newcommand{\haverford}[0]{12}
\newcommand{\mpa}[0]{13}
\newcommand{\triesteA}[0]{14}

\author{\small
D.~J.~Watts\inst{\oslo}\thanks{Corresponding author: D.~J.~Watts; \url{duncanwa@astro.uio.no}}
\and
M.~Galloway\inst{\oslo}
\and
\textcolor{black}{H.~T.~Ihle}\inst{\oslo}
\and
%\textcolor{black}{Z.~Xu}\inst{\MIT}
%\and
K.~J.~Andersen\inst{\oslo}
\and
\textcolor{black}{R.~Aurlien}\inst{\oslo}
\and
\textcolor{black}{R.~Banerji}\inst{\oslo}
% Cosmoglobe
\and
A.~Basyrov\inst{\oslo}
\and
M.~Bersanelli\inst{\milanoA, \milanoB, \milanoC}
\and
S.~Bertocco\inst{\triesteB}
\and
M.~Brilenkov\inst{\oslo}
\and
M.~Carbone\inst{\planetek}
\and
L.~P.~L.~Colombo\inst{\milanoA}
\and
H.~K.~Eriksen\inst{\oslo}
% Cosmoglobe
\and
J.~R.~Eskilt\inst{\oslo}
\and
\textcolor{black}{M.~K.~Foss}\inst{\oslo}
\and
C.~Franceschet\inst{\milanoA,\milanoC}
\and
\textcolor{black}{U.~Fuskeland}\inst{\oslo}
\and
S.~Galeotta\inst{\triesteB}
\and
S.~Gerakakis\inst{\planetek}
\and
E.~Gjerl{\o}w\inst{\oslo}
\and
\textcolor{black}{B.~Hensley}\inst{\princeton}
\and
\textcolor{black}{D.~Herman}\inst{\oslo}
\and
M.~Iacobellis\inst{\planetek}
\and
M.~Ieronymaki\inst{\planetek}
\and
J.~B.~Jewell\inst{\jpl}
\and
\textcolor{black}{A.~Karakci}\inst{\oslo}
\and
E.~Keih\"{a}nen\inst{\helsinkiA, \helsinkiB}
\and
R.~Keskitalo\inst{\nersc}
% Cosmoglobe
\and
J.~G.~S.~Lunde\inst{\oslo}
\and
G.~Maggio\inst{\triesteB}
\and
D.~Maino\inst{\milanoA, \milanoB, \milanoC}
\and
M.~Maris\inst{\triesteB}
\and
S.~Paradiso\inst{\milanoA, \milanoC}
\and
B.~Partridge\inst{\haverford}
\and
M.~Reinecke\inst{\mpa}
% Cosmoglobe
\and
M.~San\inst{\oslo}
% Cosmoglobe
\and
N.-O.~Stutzer\inst{\oslo}
\and
A.-S.~Suur-Uski\inst{\helsinkiA, \helsinkiB}
\and
T.~L.~Svalheim\inst{\oslo}
\and
D.~Tavagnacco\inst{\triesteB, \triesteA}
\and
H.~Thommesen\inst{\oslo}
\and
I.~K.~Wehus\inst{\oslo}
\and
A.~Zacchei\inst{\triesteB}
}
\institute{\small
Institute of Theoretical Astrophysics, University of Oslo, Blindern, Oslo, Norway\goodbreak
\and
%MIT Kavli Institute, Massachusetts Institute of Technology, 77 Massachusetts Avenue, Cambridge, MA 02139, U.S.A.\goodbreak
%\and
Dipartimento di Fisica, Universit\`{a} degli Studi di Milano, Via Celoria, 16, Milano, Italy\goodbreak
\and
INAF/IASF Milano, Via E. Bassini 15, Milano, Italy\goodbreak
\and
INFN, Sezione di Milano, Via Celoria 16, Milano, Italy\goodbreak
\and
INAF - Osservatorio Astronomico di Trieste, Via G.B. Tiepolo 11, Trieste, Italy\goodbreak
\and
Planetek Hellas, Leoforos Kifisias 44, Marousi 151 25, Greece\goodbreak
\and
Department of Astrophysical Sciences, Princeton University, Princeton, NJ 08544,
U.S.A.\goodbreak
\and
Jet Propulsion Laboratory, California Institute of Technology, 4800 Oak Grove Drive, Pasadena, California, U.S.A.\goodbreak
\and
Department of Physics, Gustaf H\"{a}llstr\"{o}min katu 2, University of Helsinki, Helsinki, Finland\goodbreak
\and
Helsinki Institute of Physics, Gustaf H\"{a}llstr\"{o}min katu 2, University of Helsinki, Helsinki, Finland\goodbreak
\and
Computational Cosmology Center, Lawrence Berkeley National Laboratory, Berkeley, California, U.S.A.\goodbreak
\and
Haverford College Astronomy Department, 370 Lancaster Avenue,
Haverford, Pennsylvania, U.S.A.\goodbreak
\and
Max-Planck-Institut f\"{u}r Astrophysik, Karl-Schwarzschild-Str. 1, 85741 Garching, Germany\goodbreak
\and
Dipartimento di Fisica, Universit\`{a} degli Studi di Trieste, via A. Valerio 2, Trieste, Italy\goodbreak
}

%\authorrunning{From BeyondPlanck to Cosmoglobe}
\authorrunning{Watts et al.}
\titlerunning{\cosmoglobe\ \wmap{} \Q-band analysis}

\abstract{ We present the first application of the \textsc{Cosmoglobe} analysis
framework by analyzing 9-year \textit{WMAP} time-ordered observations using
similar machinery as \textsc{BeyondPlanck} utilizes for \textit{Planck} LFI. We
analyze only the \textit Q-band (41\,GHz) data and report on the low-level
analysis process from uncalibrated time-ordered data to calibrated maps. Most
of the existing \textsc{BeyondPlanck} pipeline may be reused for \textit{WMAP}\
analysis with minimal changes to the existing codebase. The main modification
is the implementation of the same preconditioned biconjugate gradient mapmaker
used by the \textit{WMAP} team.  Producing a single \textit{WMAP} \textit
Q1-band sample requires 22\,CPU-hrs, which is slightly more than the cost of a
\textit{Planck} 44\,GHz sample of 17\,CPU-hrs; this demonstrates that full
end-to-end Bayesian processing of the \textit{WMAP} data is computationally
feasible. In general, our recovered maps are very similar to the maps released
by the \textit{WMAP} team, although with two notable differences. In
temperature we find a $\sim2\,\mathrm{\mu K}$ quadrupole difference that most
likely is caused by different gain modeling, while in polarization we find a
distinct $2.5\,\mathrm{\mu K}$ signal that has been previously called
poorly-measured modes by the \textit{WMAP} team. In the \textsc{Cosmoglobe}
processing, this pattern arises from temperature-to-polarization leakage from
the coupling between the CMB Solar dipole, transmission imbalance, and
sidelobes. No traces of this pattern are found in either the frequency map or
TOD residual map, suggesting that the current processing has succeeded in
modelling these poorly measured modes within the assumed parametric model by
using \Planck\ information to break the sky-synchronous degeneracies inherent
in the \WMAP\ scanning strategy.  }


\maketitle

\tableofcontents

\section{Introduction}
\label{sec:introduction}

Since the discovery of the cosmic microwave background (CMB;
\citealp{penzias:1965}), there have been three generations of groundbreaking
satellite missions to characterize the spatial and frequency properties of the
microwave sky; the \textit{Cosmic Background Explorer} (\textit{COBE};
\citealp{smoot:1992,mather:1994}), the \textit{Wilkinson Microwave Anisotropy
Probe} (\WMAP; \citealp{bennett2012}), and \Planck\ \citep{planck2016-l01}.
Current and future experiments designed to detect primordial gravitational
waves due to inflation \citep[e.g.,][and references therein]{kamionkowski:2016}
are built upon the foundation of these satellite missions.

The field of CMB cosmology has generally followed a model in which the data
from previous experiments are first complemented, then gradually improved upon
and superseded by those that follow. One example is \textit{COBE}/DMR, which
operated between 1989 and 1994, and discovered primordial CMB anisotropies at
31.5, 53, and 90\,GHz with a resolution of $7\degr$ \citep{smoot:1992}.
Together with the \COBE/FIRAS measurement of the CMB blackbody spectrum
\citep{mather:1994}, these observations led to the Nobel Prize in Physics in
2006. While \COBE/DMR observations were groundbreaking in their time, they have
rarely been directly used for cosmological analysis after the \WMAP\ team
released their sky maps in 2003 \citep{bennett2003a}, which improved on DMR in
terms of angular resolution and sensitivity by orders of magnitude.  When the
\Planck\ mission released its sky maps in 2013 \citep{planck2013-p01}, they
included higher sensitivity, finer angular resolution, and wider frequency
coverage, providing tighter constraints on both cosmological parameters and
Galactic physics.

Established datasets remain crucial in the analysis of current and future
datasets, both for calibration and testing, but also for breaking degeneracies
beyond their original primary science goals.  One prominent example is the
\WMAP\ \K-band sky map at 23\,GHz \citep{bennett2012}, which even after
\Planck\ represents the highest signal-to-noise ratio tracer of polarized
synchrotron emission, and is therefore used extensively for foreground
modeling. More generally, the five \WMAP\ frequencies provide essential
constraining power for low-frequency CMB foregrounds, and it is only through
the combination of \Planck\ and \WMAP\ observations (and other datasets) that
it is possible to individually constrain the properties of synchrotron,
free-free, and anomalous microwave emission (AME) over the full sky
\citep[e.g.,][]{planck2014-a12,bp13}.

A second important example is \textit{COBE}/FIRAS \citep{mather:1994}, which
showed that the CMB is well-described by the Planck blackbody radiation law at
a temperature of ${T=2.72548\pm0.00057\,\mathrm K}$ \citep{fixsen2009}; despite
being now more than 20 years old, this experiment has not yet been improved
upon, and all later CMB experiments rely directly on this value as a strong
prior for calibration purposes
\citep[e.g.,][]{bennett2012,planck2016-l05,bp07}. A third example is the
\textit{COBE}/DIRBE experiment \citep{hauser1998}, which still represents the
state-of-the-art in terms of submillimeter zodiacal light observations
\citep{kelsall1998,planck2013-pip88}, due to its unique combination of
frequency and sky coverage.  A final example is the 408\,MHz map of
\citet{haslam1982}, which is widely used as a template for Galactic synchrotron
emission. There have been attempts to improve its quality, e.g.,
\citet{remazeilles2014}, and despite its noise properties not being fully
characterized, this map is still widely used in foreground studies.

Each of these datasets faces major challenges with regard to how systematic
error correction and uncertainty propagation are handled.  In most cases, data
are provided to the public in the form of processed high-level products (most
typically pixelized sky maps, angular power spectra, or cosmological
parameters), and at these levels it is difficult to assess the impact of
instrumental effects such as calibration, beam and sidelobe errors, and
correlated noise. This in turn limits the usefulness of older datasets, as the
systematic error requirements of next-generation experiments are more stringent
than those of previous generations. A major concern is whether direct joint
analyses between old and new datasets may contaminate the latter. For almost
any new experiment, there is a tension between the desire of including
complementary datasets to break degeneracies to which one's own experiment is
not sensitive, versus the concern of introducing uncontrolled systematics into
the analysis.

\cosmoglobe\footnote{\url{https://cosmoglobe.uio.no}} aims to solve this problem by developing a common analysis platform
that is applicable to a wide range of radio, microwave, and submillimeter
experiments; legacy, current, and future. Joint analysis
of complementary experiments is essential in order to break
instrumental and astrophysical parameter degeneracies.  Therefore, as
more datasets are added to this analysis, better cosmological and
astrophysical results will emerge. Enabling and organizing this
work is the main goal of the community-wide and Open Science
\textsc{Cosmoglobe} program.

\BP\ represents the first stage of the process, in which the \Planck\ Low
Frequency Instrument (LFI;
\citealp{planck2013-p03,planck2014-a03,planck2016-l02}) data are processed within a global Bayesian framework. This data set was chosen for three reasons; 1) the LFI data volume is relatively low, allowing
for fast debugging; 2) the LFI instrumental systematics are well understood;
and 3) the current team members have years of experience working with this
dataset.

In this paper, 
we generalize the same framework to support \WMAP\ time domain analysis. We note that a full \WMAP\ reanalysis lies outside
the scope of the present paper, as this will require both additional modeling
and analysis effort. Rather, the main goals of the current work are to answer
the following practical questions: First, how much software development effort
is required to generalize the \commander\ software to support an entirely new
dataset? Second, is a proper end-to-end Bayesian analysis of the full \WMAP\
dataset technically feasible with currently available computing power, and if
so, how much computational power will it require? Third, what additional
instrument-specific features are required to perform a full \WMAP\ analysis?
These questions are important not only for a future \WMAP\ reanalysis itself,
but also for other experiments considering adopting the \commander\ framework for
their own analysis. In implementing a \WMAP\ pipeline using \commander, we
demonstrate that this Bayesian framework can be generalized to datasets beyond the one it was
explicitly designed to analyze, \Planck\ LFI. 

\section{Bayesian \WMAP\ analysis with \commander}
\label{sec:bp}

\subsection{The \BP\ data model and posterior distribution}

The \BP\ framework \citep{bp01} takes a novel approach to CMB data analysis by
adopting a strictly parametric Bayesian end-to-end formulation. As for any
parametric Bayesian calculation, the first step in implementing the algorithm
is writing down an explicit parametric data model; everything else will, ideally,
follow naturally from that model. The data model adopted for \Planck\
LFI takes the form of
\begin{equation} d_{t,j}=g_{t,j} \mathsf P_{tp,j}\left[
	\mathsf B_{j}^\mathrm{mb} s_{j}^\mathrm{sky} + \mathsf
	B_{j}^\mathrm{fsl} s_{j}^{\mathrm{sky}} +\mathsf
	B_{j}^\mathrm{4\pi}s_{t,j}^\mathrm{orb} \right]
	+ s_{t,j}^{\mathrm{1\,Hz}}+n_{t,j}^\mathrm{corr}+n_{t,j}^\mathrm w,
	    \label{eq:todmodel}
\end{equation}
where $t$ indexes the observation time, $j$ indexes the
detector, $p$ indexes the pixel, $g_{t,j}$ is a time-dependent gain, $\mathsf P_{tp,j}$ is a pointing
matrix, $\mathsf B$ includes components of the beam (main beam $\tens
B^\mathrm{mb}$, far sidelobes $\tens B^\mathrm{fsl}$, and full $\tens
B^\mathrm{4\pi}$, respectively), $s_{j}^\mathrm{sky}$ is the (time-independent)
sky signal, $s_{t,j}^\mathrm{orb}$ is the orbital CMB dipole,
${s_{t,j}^\mathrm{fsl}=\mathsf B_j^\mathrm{fsl}s_{j}^\mathrm{sky}}$ is the
(orientation-dependent) far sidelobe contribution,
$s_{t,j}^\mathrm{1\,Hz}$ is a contribution from electronic 1\,Hz spikes, $n_{t,j}^\mathrm{corr}$ is
the correlated noise, and $n_{t,j}^\mathrm w$ is the white noise. Note that the 1\,Hz spikes are unique to \Planck\ LFI, and are not used in the \wmap\ analysis.
The parametric sky model $\boldsymbol s^\mathrm{sky}(\nu,\boldsymbol a,\boldsymbol\beta)$ includes contributions from CMB, synchrotron, free-free, spinning dust, thermal dust, and point source emission. A full description of the sky model can be found in \citet{bp01}.

The data model may be written
in a compact vector form
\begin{equation}
	\boldsymbol d=\mathsf{GPBM}\boldsymbol a+\s^{\mathrm{fsl}} +
        \s^{\mathrm{orb}} + \boldsymbol n \equiv \s^{\mathrm{tot}} + \n, 
        \label{eq:vecmodel}
\end{equation}
where we have now also introduced the diagonal gain matrix ${\mathsf G_j=\mathrm{diag}(g_{t,j})}$ and the mixing matrix $\mathsf M$ to
describe bandpass integration effects,
\begin{equation}
	\mathsf M_j^i\equiv \int
	f_i(\nu;\boldsymbol\beta)\,U_j(\Delta_\mathrm{bp})\,\tau_j(\nu; \Delta_{\mathrm{bp}})\,\mathrm d\nu.
        \label{eq:mixmat}
\end{equation}
Here $\tau_j$ is the bandpass for each detector with a free parameter
$\Delta_{\mathrm{bp}}$, $U_j$ converts from brightness temperature to frequency
sky map unit integrated over the bandpass \citep{planck2013-p03d},
$f_i(\nu;\boldsymbol\beta)$ is the SED of component $i$ given the generalized
SED parameters $\boldsymbol\beta$, and ${\n = \n^{\mathrm{corr}} +
\n^{\mathrm{wn}}}$. For LFI, $\n$ is often assumed to be Gaussian distributed
with a covariance matrix, $\N$, given by a $1/f$ power spectral density (PSD), 
\begin{equation}
	\mathcal P(f) =
\sigma^2[1+(f/f_{\mathrm{knee}})^\alpha],
	\label{eq:psd}
\end{equation}
where $\sigma$ is the white noise standard deviation, $f_{\mathrm{knee}}$ is
the correlated noise knee frequency, and $\alpha$ is the correlated noise
spectral index. In general, we will denote the set of all noise PSD parameters
as $\boldsymbol\xi_n$. For a full discussion of
this model, we refer the interested reader to \citet{bp01} and references
therein.

Let us now denote the set of all free parameters in
Eqs.~\eqref{eq:todmodel}--\eqref{eq:mixmat} by $\boldsymbol \omega$, such that
$\boldsymbol \omega = \{\g, \ncorr, \boldsymbol\beta, \a, \ldots\}$. The Bayesian approach is
to map out the posterior distribution, 
\begin{equation}\label{eq:full_distribution}
	P(\boldsymbol \omega\mid\d) \propto \mathcal{L}(\d\mid\boldsymbol \omega) P(\boldsymbol \omega),
\end{equation}
using standard Markov Chain Monte Carlo sampling methods, where
$\mathcal{L}(\boldsymbol \omega)$ is the likelihood function, and
$P(\boldsymbol \omega)$ is the prior probability of the vector $\boldsymbol
\omega$. We define the likelihood by assuming that the noise component in
Eq.~\eqref{eq:vecmodel} is Gaussian distributed, such that
\begin{equation}
	-2\ln\mathcal{L}(\boldsymbol \omega) = (\d-\s^{\mathrm{tot}}(\boldsymbol \omega))^T\N^{-1}(\boldsymbol \omega)(\d-\s^{\mathrm{tot}}(\boldsymbol \omega)) + \ln|2\pi\,\mathsf N|.
\end{equation}
The priors are in general less well-defined, and in practice we use both
algorithmic and informative priors to ensure a robust fit; see, for example,
\citet{bp01} and \citet{bp13}. For the special case of the CMB, it is common to
assume that its fluctuations are isotropic and Gaussian distributed, with a
variance given by the angular power spectrum, $C_{\ell}$; estimating this power
spectrum is typically a main goal for most CMB experiments.

The posterior distribution defined by Eq.~\eqref{eq:full_distribution} is
infeasible to map out directly, due to the sheer number of free parameters and
degeneracies within the model. However, the Gibbs sampling algorithm
\citep{gelman:1992} allows for efficient exploration of the full joint
distribution by iterating through all conditional distributions, each of
which are simpler to explore than the full joint distribution. To be specific,
the \BP\ Gibbs chain takes the form \citep{bp01},
\begin{alignat}{11}
\label{eq:gain_samp_dist}\g &\,\leftarrow          P(\g&\,               \mid \d, &\,    &          &\,\boldsymbol\xi_n,  &\,\Dbp, &\,\boldsymbol\beta, &\,\a, &\,C_{\ell})\\
\label{eq:ncorr_samp_dist} \ncorr &\,\leftarrow    P(\ncorr&\,           \mid \d, &\,\g, &\,        &\,\boldsymbol\xi_n,  &\,\Dbp, &\,\boldsymbol\beta, &\,\a, &\,C_{\ell})\\ 
\label{eq:xi_samp_dist} \boldsymbol\xi_n &\,\leftarrow        P(\boldsymbol\xi_n&\,            \mid \d, &\,\g, &\,\ncorr, &\,        &\,\Dbp, &\,\boldsymbol\beta, &\,\a, &\,C_{\ell})\\
\Dbp &\,\leftarrow                                 P(\Dbp&\,             \mid \d, &\,\g, &\,\ncorr, &\,\boldsymbol\xi_n,  &\,      &\,\boldsymbol\beta, &\,\a, &\,C_{\ell})\\
\boldsymbol\beta &\,\leftarrow                     P(\boldsymbol\beta&\, \mid \d, &\,\g, &\,\ncorr, &\,\boldsymbol\xi_n,  &\,\Dbp, &\,       &\,    &\,C_{\ell})\\
\a &\,\leftarrow                                   P(\a&\,               \mid \d, &\,\g, &\,\ncorr, &\,\boldsymbol\xi_n,  &\,\Dbp, &\,\boldsymbol\beta, &\,    &\,C_{\ell})\\
C_{\ell} &\,\leftarrow                             P(C_{\ell}&\,         \mid \d, &\,\g, &\,\ncorr, &\,\boldsymbol\xi_n,  &\,\Dbp, &\,\boldsymbol\beta, &\,\a \phantom{,}&\,\phantom{C_{\ell}})&\label{eq:cl_sampling},
\end{alignat}
where $\leftarrow$ denotes drawing a sample from the conditional
distribution on the right. In the \commander\ framework, CMB analysis essentially
amounts to repeating each of these steps until convergence, which typically
requires thousands of iterations. 

In this work, we hold the amplitude and SED parameters $\boldsymbol a$
and $\boldsymbol\beta$ fixed, as our goal is to determine the
feasiblility of extending the time-ordered data (TOD) analysis to the
\WMAP\ dataset. In a full analysis, we would perform a full Gibbs
chain on all of these parameters jointly, but in this paper we first
fit the sky model using almost the same data combination as
\citet{bp01}, i.e., \Planck\ LFI 30--70\,GHz, \Planck\ HFI 353 (in
polarization) and 857\,GHz (in temperature), \WMAP9 \Ka--\V, and Haslam
408\,MHz. However, unlike the main analysis, we additionally include
the \WMAP9 \textit K-band data to increase the signal-to-noise ratio
for low-frequency foreground components.  We then use this fixed sky
model throughout to calibrate the \Q1 data.
This yields a simplified Gibbs chain,
\begin{alignat}{11}
\g &\,\leftarrow          P(\g&\,               \mid \d, &\,    &          &\,\boldsymbol\xi_n,  &\,\Dbp, &\,\boldsymbol\beta, &\,\a, &\,C_{\ell})\\
\ncorr &\,\leftarrow    P(\ncorr&\,           \mid \d, &\,\g, &\,        &\,\boldsymbol\xi_n,  &\,\Dbp, &\,\boldsymbol\beta, &\,\a, &\,C_{\ell})\\ 
\boldsymbol\xi_n &\,\leftarrow        P(\boldsymbol\xi_n&\,            \mid \d, &\,\g, &\,\ncorr, &\,        &\,\Dbp, &\,\boldsymbol\beta, &\,\a, &\,C_{\ell}),
\end{alignat}
which holds all sky and bandpass parameters fixed throughout.

\subsection{Generalization to \WMAP}

The \WMAP\ mission \citep{bennett2012} observed the sky at \K, \Ka, \Q, \V, and
\W-bands (23, 33, 41, 61, and 94\,GHz, respectively) using differential
radiometers, and observed from August~10, 2001 to August~10, 2010. The
satellite observed from the second Sun-Earth Lagrange point with a Lissajous
orbit, rotating around its primary axis with a period of 129\,s, and precessing
around its spin axis with a period of one hour, allowing for total coverage of
the sky every six months. This observing strategy allowed for excellent control
of systematic effects that appear in the time streams \citep{bennett2003:MAP}.

The \WMAP\ instrument is inherently differential, with each detector recording
the signal difference received in two horns, labeled A and B. Each radiometer
is sensitive to two orthogonal polarization directions, $\gamma$ and
$\gamma+\pi/2$, and each of these are coupled pairwise with the two horns,
such that each radiometer ultimately results in
four polarized time streams, which are treated jointly in a  single
differencing assembly (DA). 
\WMAP\ has ten total DAs, one for both \K\ and \Ka, two each for \Q\ and \V, and four for \W.
In this paper, we consider the DA corresponding to
\Q1, whose individual time streams are indexed by \Q1\textit{ij}, where
$i\in\{1,2\}$ labels the detector's polarization orientation, and $j\in\{3,4\}$
labels the $s_\A-s_\B$ and $s_\B-s_\A$ time streams.

The primary goal of the current paper is to understand what is required in
terms of coding efforts and computational resources in order to apply the \commander\
framework as summarized above to the \WMAP\ data. The first step in that
process is to write down an explicit parametric data model. When one reviews
the descriptions of the \WMAP\ instrument and time-ordered data provided by
\citet{bennett2003a}, \citet{barnes2003}, \citet{jarosik2003},
\citet{hinshaw2003a}, \citet{page2007}, \citet{jarosik2007}, and
\citet{wmapexsupp}, it becomes clear that the LFI data model defined in
Eq.~\eqref{eq:todmodel} also applies to \WMAP\ with only three modifications, 
two major and one minor. We will now address each of these in turn.

First, while \Planck\ measures the power received from a single point on the sky, \WMAP\
records the difference between two points in digital units
(du),
\begin{equation}
	\begin{split}
	s=%\quad
		g\big[\,&\alpha_\A(T_\A + Q_\A\cos2\gamma_\A+U_\A\sin2\gamma_\A)
		\\
		\,-&\alpha_\B(T_\B + Q_\B\cos2\gamma_\B+U_\B\sin2\gamma_\B)\big]
	\end{split}.
        \label{eq:wmap_signal0}
\end{equation}
In this equation, $\{T,Q,U\}$ are the Stokes parameters seen by each horn, and
$\alpha_{\A/\B}$ is a horn transmission coefficient that quantifies the
transmission of the optics and waveguide components, which may be slightly different
between the two sides.  Following \citet{hinshaw2003a}, the transmission
coefficients are written in terms of horn imbalance parameters
$x_\mathrm{im}$ that explicitly parameterize the difference between the A and B
horn transmissions,
\begin{equation}
	x_{\mathrm{im}}=\frac{\alpha_{\A}-\alpha_{\B}}{\alpha_{\A}+\alpha_{\B}}.
\end{equation} 
The overall normalization is absorbed into the gain, so that the data model becomes
\begin{equation}
	\begin{split}
	s=%\quad
		g\big[\,&(1+x_{\mathrm{im}})(T_\A + Q_\A\cos2\gamma_\A+U_\A\sin2\gamma_\A)
		\\
		\,-&(1-x_\mathrm{im})(T_\B + Q_\B\cos2\gamma_\B+U_\B\sin2\gamma_\B)\big]
	\end{split}.
        \label{eq:wmap_signal}
\end{equation}

Equation~\eqref{eq:wmap_signal} may now be implemented in the LFI data
model in Eq.~\eqref{eq:todmodel} by redefining the pointing
matrix, such that a single row reads
\begin{equation}
  \mathsf P_{tp}=
  \left[
    \begin{array}{c}
      0\\
      \vdots\\
      0\\
      (1+x_{\mathrm{im}})\phantom{\cos2\gamma_\A}\\
      (1+x_{\mathrm{im}})\cos2\gamma_\A\\
      (1+x_{\mathrm{im}})\sin2\gamma_\A\\
      0\\
      \vdots\\
      0\\
	  -(1-x_{\mathrm{im}})\phantom{\cos2\gamma_\A}\\
          -(1-x_{\mathrm{im}})\cos2\gamma_\B\\
          -(1-x_{\mathrm{im}})\sin2\gamma_\B\\
          0\\
          \vdots\\
          0\\
    \end{array}
    \right]^T  
\end{equation}
Therefore, the fact that \WMAP\ records differential pointing while \Planck\
records the signal from a single point in the sky implies that the \WMAP\
pointing matrix has twice as many entries as for LFI, and that there is also an
additional uncertain parameter per radiometer that needs to be sampled for
\WMAP, the transmission imbalance parameter. In terms of implementation, this
also means that the most important generalization required for Bayesian
analysis of \WMAP\ is the implementation of a mapmaker for differential data. This
task has already been addressed by the \WMAP\ team, who showed that a
stabilized biconjugate gradient method works well for this problem
\citep{jarosik2010}. The main recoding effort done in this paper is thus to
reimplement this method in \commander, as discussed in
Sect.~\ref{sec:mapmaking}. 

The second main difference between \WMAP\ and LFI as defined by
Eq.~\eqref{eq:todmodel} lies in the noise model. While the LFI noise
is close to white at high temporal frequencies and can be described
well with a $1/f$ model (or, at least, as a sum of a $1/f$ model and a
subdominant Gaussian peak; see \citealp{bp06}), the \WMAP\ noise is
in general colored close to the Nyquist frequency, typically
exhibiting a very slight power increase at the highest
frequencies. This does have some important sampling technical
implications for the noise PSD Gibbs step, as defined in
Eq.~\eqref{eq:xi_samp_dist} and discussed by \citet{bp06}. The current
LFI implementation assumes that the noise is white at the sampling
frequency, and explicitly uses this to break a degeneracy between the
correlated and white noise components.  At the same time, the amplitude of this
\WMAP\ colored high-frequency noise is very modest, and as shown in
Sect.~\ref{sec:Results} a standard $1/f$ model does fit reasonably
well. Furthermore, a suboptimal noise model will result in
slightly suboptimal uncertainties, but not biases.  Therefore, this is
a minor issue for the current paper, which aims to assess the overall
applicability of the Bayesian approach for \WMAP; low-level noise
modeling issues do not affect this question. We also note that we need
to generalize the current noise model to allow for analysis of
\Planck\ HFI and other bolometer experiments. An expansion of the
\commander{} noise model is therefore left for future work.

The third and final difference between LFI and \WMAP\ are electronic
1\,Hz spikes, which are not relevant for \WMAP. This term is therefore
omitted in the following \WMAP\ analysis.

\begin{figure}
	\includegraphics[width=\columnwidth]{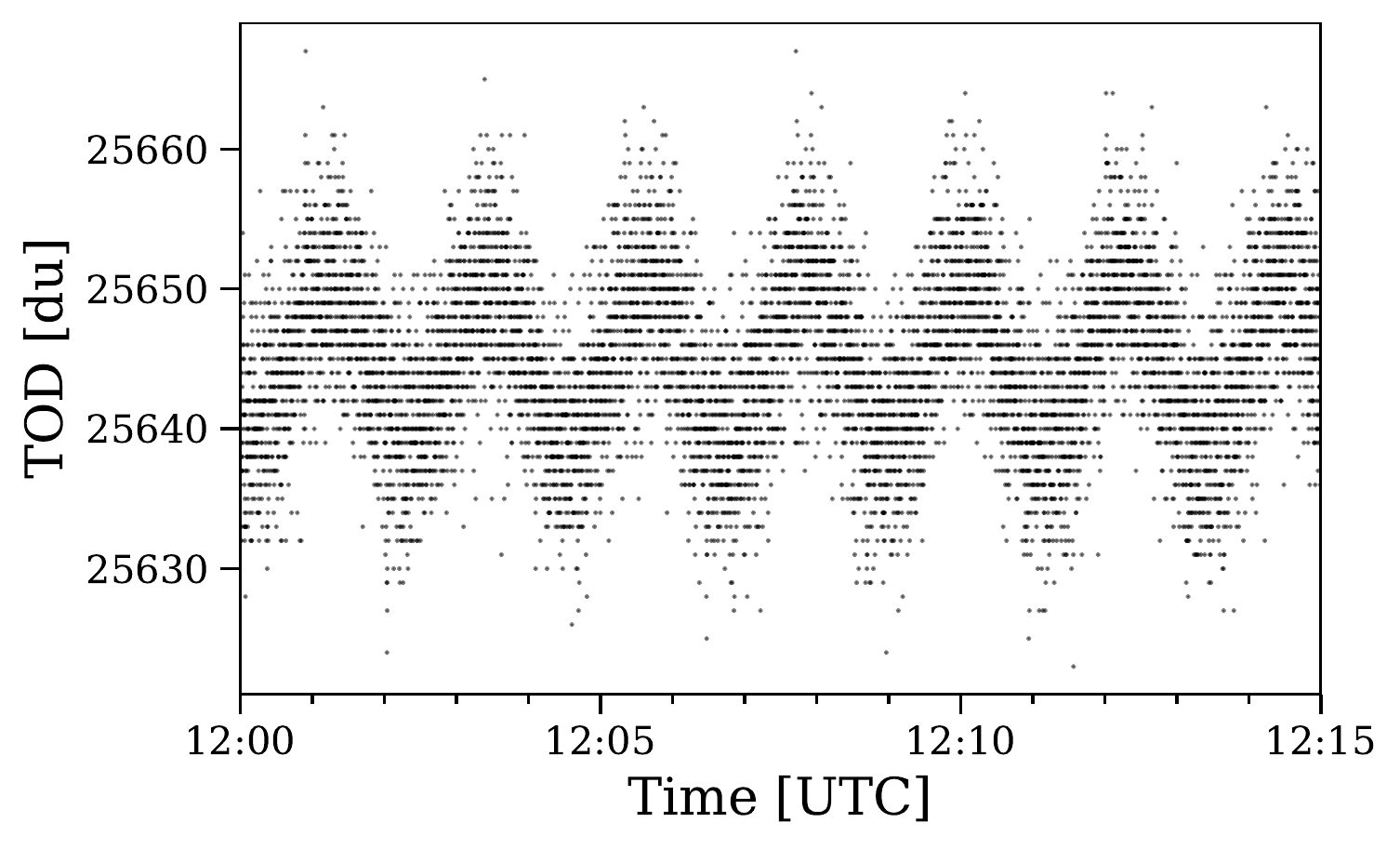}
	\caption{Sample of \WMAP\ TOD. This \Q113 datastream was recorded on July 7, 2003. At this resolution, the discreteness of the digital units is apparent, a property which makes these data highly compressible.}
	\label{fig:tod}
\end{figure}

All other parameters and sampling steps in
Eqs.~\eqref{eq:todmodel}--\eqref{eq:cl_sampling} are identical between
LFI and \WMAP. Since the two experiments cover roughly the same
frequency range and angular scales, no new astrophysical components need to be added to
the sky model (with the possible exception of HCN and other line
emission at \W-band, as discussed by \citealp{planck2014-a12}), and
the required low-level algorithms for sidelobe convolution and orbital
dipole generation are identical between the two experiments. Thus,
\WMAP\ is an excellent example of the benefits of joint analysis
within a single computational framework; nearly all of the existing
computer code is directly reusable.

\section{Algorithmic details}
\label{sec:algorithms}

The full 9-year \WMAP\ dataset\footnote{\url{https://lambda.gsfc.nasa.gov/product/map/dr5/tod_uncal_get.cfm}}
spans 626\,GB, which represented a major challenge with respect to optimal
mapmaking using all available data in 2013, simply due to computer hardware
limitations. In practice, this was overcome by processing each year of
observations separately, and then creating a noise-weighted average of the
maps. During mapmaking, the TOD were processed in one~hour or daily chunks
\citep{bennett2012}.

Today, we circumvent these data volume issues two different ways within \commander. First, having access to large-memory compute nodes with 1.5\,TB
of RAM greatly alleviates hardware based concerns. This is further improved by storing the TOD in a compressed format in RAM, not
only on disk. Specifically, as described by \citet{bp03}, the current pipeline
uses Huffman compression to store the TOD, which reduces the number of bits per
stored number according to the frequency of that same number. This technique is
particularly powerful for the \WMAP\ data, as illustrated in
Fig.~\ref{fig:tod}. The most striking feature of this data stream is its
discreteness, imposed by the analog-to-digital converter; while the data are
delivered in terms of 32\nobreakdash-bit integers, typically only 50 of those are
encountered in any given data segment. Therefore, relabeling the relevant
integers with shorter bit strings yields a significant reduction in data volume. We
find that the entire \WMAP\ TOD only requires 186\,GB of disk storage after Huffman
compression, at which point they may be stored in RAM even on inexpensive modern
compute nodes.\footnote{\href{http://sdc.uio.no/vol/cosmoglobe-data/WMAP/TODs/}
{\texttt{http://sdc.uio.no/vol/cosmoglobe-data/WMAP/TODs/}}}

While \WMAP\ used either one or 24~hour chunk sizes for their TOD processing
\citep{bennett2012}, we adopt one~week periods for our processing. This choice
is informed by the extremely low levels of correlated noise in the \WMAP\ data,
with $f_{\mathrm{knee}}\lesssim 1\,\mathrm{mHz}$ for half of the radiometers
\citep{jarosik2003}, corresponding to 20 minutes or more in the time domain.
With one~week time chunks, it is easier to disentangle white and correlated
noise, and the processing is less sensitive to Fourier-filtering edge effects
and aliasing.

In this section, we consider in greater detail the three specific algorithmic
changes that are necessary for \WMAP\ processing within the \commander\ framework,
namely 1) transmission imbalance sampling; 2) mapmaking with differential data;
and 3) noise modeling, which are all summarized in Sects.~\ref{subsec:gainimbal}--\ref{sec:psd}. In addition, we review the \commander\ approach to sidelobe
corrections in Sect.~\ref{sec:sidelobes}. We summarize the algorithmic differences between \wmap\ and \cosmoglobe\ in Sect.~\ref{sec:differences}.

\subsection{Gain and Imbalance Sampling}
\label{subsec:gainimbal}

For the purposes of gain sampling, as symbolically defined by
Eq.~\eqref{eq:gain_samp_dist}, we can simplify the global parametric
model in Eq.~\eqref{eq:todmodel} to 
\begin{equation}
	d_{t,j} = g_{t,j}^{\phantom{tot}}s_{t,j}^\mathrm{tot}+n_{t,j}^\mathrm{corr}+
	n_{t,j}^\mathrm{wn},
\end{equation}
where $s_{t,j}^\mathrm{tot}$ is the full beam-convolved sky signal in
time domain for radiometer $j$. For \WMAP, this model may be
generalized to differential data with imbalance parameters
$x_\mathrm{im}$ as
\begin{equation}
	d_{t,j} = g_{t,j}[(1+x_{\mathrm{im},j})s_{t,j}^\mathrm{tot,A}
			 - (1-x_{\mathrm{im},j})s_{t,j}^\mathrm{tot,B}]
	+n_{t,j}^\mathrm{corr}+
	n_{t,j}^\mathrm{wn}.
       \label{eq:wmap_gain}
\end{equation}
To sample $g_{t,j}$, we adopt the standard \BP\ procedure without modification,
and define $g_{t,j}=g_0+\Delta g_j+\delta g_{t,j}$. Here, $g_0$ denotes the
time-independent absolute calibration for the entire DA, $\Delta g_j$
represents the time-independent offset from $g_0$ for radiometer $j$, and
$\delta g_{t,j}$ represents the time-dependent fluctuations around the mean for
radiometer $j$. Each of these three terms is sampled conditionally using its
own tuned algorithm. Specifically, $g_0$ is sampled using the orbital CMB
dipole only as a calibration source, $\Delta g_j$ is sampled using the full
astrophysical sky model, including the Solar dipole, and $\delta g_{t,j}$ is
sampled using an optimal Wiener filter algorithm that weights time-variable
fluctuations according to their relative signal-to-noise ratio. For full
details, we refer the interested reader to \citet{bp07}.

From Eq.~\eqref{eq:wmap_gain}, we see that $x_{\mathrm{im},j}$ plays a role that is
similar to $g_0+\Delta g_j$. However, unlike those parameters, it applies to
pairs of time streams, and for the \Q1 DA we estimate the transmission
imbalance for \Q11 using time streams \Q113 and \Q114, and the transmission
imbalance for \Q12 using time streams \Q123 and \Q124.

To sample $x_\mathrm{im}$, we follow \citet{hinshaw2003a} and define
the following differential (d) and common-mode (c) signals,
\begin{align}
	\vec s^\mathrm{d}&=\vec s^\mathrm{tot,A}-\vec s^\mathrm{tot,B}
	\\
	\vec s^\mathrm{c}&=\vec s^\mathrm{tot,A}+\vec s^\mathrm{tot,B},
\end{align}
such that
\begin{equation}
	d_{t,j}=g_{t,j}[s_{t,j}^\mathrm d
	+x_{\mathrm{im},j}^{\phantom{c}}s_{t,j}^\mathrm c]
	+n_{t,j}^\mathrm{corr}+n_{t,j}^\mathrm{wn}.
        \label{eq:data_tod}
\end{equation}
If we now form the residual $r_{t,j}\equiv
d_{t,j}-g_{t,j}^{\phantom{d}}s_{t,j}^\mathrm d$ and define ${\mathsf T\equiv
g_{t,j}^{\phantom{c}}s_{t,j}^\mathrm{c}}$, the equation for the residual becomes
\begin{equation}
	\r=\T \boldsymbol x_{\mathrm{im}}
	+\n^\mathrm{corr}+\n^\mathrm{wn}.
\end{equation}
Since we assume the noise is Gaussian distributed with covariance
$\N$, and both $\T$ and $\s^{\mathrm{c}}$ are conditionally fixed, 
the appropriate conditional distribution for $x_{\mathrm{im},j}$
is also a Gaussian. We can therefore follow the standard procedure
outlined in Appendix~A.2 of
\citet{bp01}, and sample $x_\mathrm{im}$ through the following equation,
\begin{equation}
	(\mathsf T^T\mathsf N^{-1}\mathsf T)\boldsymbol x_\mathrm{im}
	=\mathsf T^T\mathsf N^{-1}\boldsymbol r
	+\mathsf T^T\mathsf N^{-1/2}\boldsymbol\eta,
\end{equation}
where $\boldsymbol \eta \sim N(\boldsymbol 0,\mathsf I)$ is a vector of standard Gaussian random
variates. This may be written explicitly as
\begin{equation}
  \label{eq:imbalsamp}
  x_{\mathrm{im},j}=
  \frac{\sum_{i,t,t'}[g_{i,t}s_{i,c,t}^\mathrm{c}]
  \mathsf N^{-1}_{tt'}r_{i,t'}}
  {\sum_{i,t,t'}[g_{i,t}s_{i,t,c}^\mathrm{c}]
  \mathsf N^{-1}_{tt'}[g_{i,t'}s_{i,t',c}^\mathrm{c}]}
  +
  \frac{\sum_{i,t,t'}[g_{i,t}s_{i,c,t}^\mathrm{c}]
  \mathsf N^{-1/2}_{tt'}\eta_{t'}}
  {\sum_{i,t,t'}[g_{i,t}s_{i,t,c}^\mathrm{c}]
  \mathsf N^{-1}_{tt'}[g_{i,t'}s_{i,t',c}^\mathrm{c}]},\nonumber
\end{equation}
where $i$ corresponds to the data streams for radiometer pair $j$.

\subsection{Mapmaking}
\label{sec:mapmaking}

While most instrumental parameters are sampled in the time domain in the
Bayesian framework, astrophysical parameters are still sampled in
terms of pixelized sky maps. We therefore need to compute
coadded maps for each DA. The starting point of this process is the
calibrated residual,
\begin{equation}
r_{t,j}=\frac{d_{t,j}-n_{t,j}^\mathrm{corr}}{g_{t,j}}-(s_{t,j}^\mathrm{orb}+s_{t,j}^\mathrm{fsl}+\delta
s_{t,j}^\mathrm{leak}),
\label{eq:map_res}
\end{equation}
where we condition on all instrumental parameters, including
$\g$, $\n^{\mathrm{corr}}$, $\boldsymbol\xi_n$, and
$\boldsymbol x_{\mathrm{im}}$. In this expression, we have also defined
\begin{equation}
	\delta s_{t,j}^\mathrm{leak}=\mathsf P_{tp,j}\mathsf B_{pp',j}\left(
	s_{jp'}^\mathrm{sky}-\left\langle
        s_{jp'}^\mathrm{sky}\right\rangle\right),
\end{equation}
which is the difference between the sky signal seen by radiometer $j$ and the
same averaged over all radiometers in the given DA. This term accounts for
spurious bandpass leakage effects without solving for a spurious map per pixel,
as discussed by \citet{bp09}.

Inserting Eq.~\eqref{eq:map_res} into the data model in
Eq.~\eqref{eq:todmodel}, we find 
\begin{align}
	r_{t,j}=&\,\mathsf P_{tp,j}s_{jp}+n_{t,j}
	\\
	=&\,(1+x_{\mathrm{im},j})[T_{p_\A}+P_{p_\A}(\gamma_{\A,t})]
	\nonumber
	\\
	&\!\!\!\!\!-(1-x_{\mathrm{im},j})[T_{p_\B}+P_{p_\B}(\gamma_{\B,t})]
	+n_{t,j},
	\label{eq:skymodel}
\end{align}
where we denote the total polarized signal as
${P_{p_{\A/\B}}(\gamma_{\A/\B,t})=Q_{p_{\A/\B,t}}\cos2\gamma_{\A/\B,t}
+U_{p_{\A/\B,t}}\sin2\gamma_{\A/\B,t}}$. Following \citet{jarosik2010},
we combine these calibrated data into an ``intensity'' time stream $d_t$
and a ``polarization'' time stream $p_t$,
\begin{align}
	d_t&=\frac14(d_{13,t}+d_{14,t}+d_{23,t}+d_{24,t})
	\\
	&=T_{p_\A}-T_{p_\B}
	+\bar x_\mathrm{im}[T_{p_\A}+T_{p_\B}]
	\nonumber
	\\
	&\,\,\,\,\,\,\,\,+\delta x_\mathrm{im}[P_{p_\A}(\gamma_\A)+P_{p_\B}(\gamma_\B)],
	\\
	p_t&=\frac14(d_{13,t}+d_{14,t}-d_{23,t}-d_{24,t})
	\\
	&=P_{p_\A}(\gamma_{\A})-P_{p_\B}(\gamma_{\B})
	+\bar x_\mathrm{im}[P_{p_\A}(\gamma_{\A})+P_{p_\B}(\gamma_{\B})]
	\nonumber
	\\
	&\,\,\,\,\,\,\,\,+\delta x_\mathrm{im}[T_{p_\A}+T_{p_\B}],
\end{align}
where $\bar x_\mathrm{im}=(x_\mathrm{im,1}+x_\mathrm{im,2})/2$ and $\delta
x_\mathrm{im}=(x_\mathrm{im,1}-x_\mathrm{im,2})/2$.  This formalism
approximately splits the data streams into intensity-only and
polarization-only, except for terms proportional to $\delta x_\mathrm{im}$,
which is a factor of $\lesssim10^{-3}$. This term must be specifically
accounted for in the polarization time stream, since $\delta
x_\mathrm{im}[T_{p_\A}+T_{p_\B}]$ has a nonnegligible amplitude compared to
$P_{p_\A}-P_{p_\B}$.

Because two pointings contribute to a single observation, we cannot
solve for the underlying sky map pixel-by-pixel as is the case with
\Planck. However, the more general maximum likelihood mapmaking equation
\begin{equation}
	\mathsf P^T\mathsf N^{-1}\mathsf P\boldsymbol m=\mathsf P^T\mathsf N^{-1}\boldsymbol d,
\end{equation}
still applies, and the only real difference is the structure of the pointing matrix $\P$,
as already discussed above.

There is an additional complication that arises due to differences in horn A
and horn B and the different pixel shapes that are being binned. Differences
between the data and the model become exacerbated when one horn is observing a
bright source and the other is not. For this reason, we follow the \WMAP{}
team's procedure of asymmetric masking. This entails using a pointing matrix
$\mathsf P_\mathrm{am}$ that masks pixel B when pixel A observes a bright spot
and pixel B is observing a relatively faint point, and vice versa. We use the
same processing mask that is used for LFI to define these bright spots. This
gives the modified mapmaking equation
\begin{equation}
	\label{eq:bicgstab}
	\mathsf P_\mathrm{am}^T\mathsf N^{-1}\mathsf P\boldsymbol m=\mathsf P_\mathrm{am}^T\mathsf N^{-1}\boldsymbol d.
\end{equation}
This form of the mapmaking equation gives an unbiased estimate of the map,
while the pixel space covariance matrix is slightly larger due to the data cuts
in the asymmetric masking. To solve this equation, we follow
\citet{jarosik2010} and use the stabilized biconjugate gradient method
(BiCG-STAB; \citealt{bicgstab,bicgstab_template}) to solve for this map. Note
that we cannot use the usual conjugate gradient algorithm because the matrix
$\mathsf P_\mathrm{am}^T\mathsf N^{-1}\mathsf P$ is not symmetric.

Solving Eq.~\eqref{eq:bicgstab} requires iterating over every observation in the TOD,
and is currently the costliest step in the \WMAP{} analysis
pipeline. This typically takes about 20 BiCG iterations, each taking
roughly 11\,s of wall time or 12\,CPU-mins, for a total of about 4\,CPU-hrs per sky map. There
are several ways to optimize this technique, including using a good
initial guess for the map, or a well-chosen preconditioner. Although
not yet implemented in our code, we note that \citet{jarosik2010}
derive the inverse pixel-pixel noise covariance
\begin{equation}
	\mathsf{\Sigma}^{-1}=(\mathsf P^T\mathsf N^{-1}\mathsf P_\mathrm{am})(\mathsf P_\mathrm{am}^T\mathsf N^{-1}\mathsf P_\mathrm{am})^{-1}(\mathsf P_\mathrm{am}^T\mathsf N^{-1}\mathsf P),
\end{equation}
and use the central term $\mathsf P_\mathrm{am}^T\mathsf N^{-1}\mathsf P_\mathrm{am}$ evaluated at $N_\mathrm{side}=16$ as a source for the off-diagonal terms in the preconditioner. 

\subsection{Noise modeling}

\label{sec:psd}

Next, we consider the temporal noise model.  The \WMAP\ radiometers have
remarkably white noise, with $f_\mathrm{knee}$ ranging from 0.1--40\,mHz, with
typical values around 1\,mHz \citep{jarosik2003}. For comparison, the LFI
time-ordered data have knee frequencies ranging from 5--200\,mHz, with typical
values in the 50\,mHz region \citep{bp06}. Such long stability periods require
more careful analysis in order to characterize the small deviations from white
noise in the \WMAP\ data.  The \WMAP\ team's approach was based on a
third-order polynomial fit to the two-point temporal correlation function
\begin{equation}
	N(\Delta t)=
	\begin{cases}
		AC & \Delta t=1,
		\\
		\displaystyle{\sum_{n=0}^3} a_n[\log_{10}(|\Delta t|)]^n & 1<|\Delta t|<\Delta t_\mathrm{max},
		\\
		0 & |\Delta t|\geq \Delta t_\mathrm{max},
	\end{cases}
	\label{eq:noise_acf}
\end{equation}
where $AC$ and $a_n$ are parameters that were fit to the autocorrelation data,
$\Delta t$ is in units of samples, and $\Delta t_\mathrm{max}$ corresponds to
the time lag where the fit crosses zero, typically $\sim600\,\mathrm s$
\citep{jarosik2007}.  We convert the parametric noise autocorrelation function\footnote{\url{https://lambda.gsfc.nasa.gov/product/wmap/dr5/tod_filters_info.html}}
to a time-domain power spectrum by taking the Fourier transform, $\mathcal P(f)=\Big[\mathcal
F[N(\Delta t)]\Big]^{-1}$. In contrast, the current LFI-based \BP\ noise model
is primarily based on a standard $1/f$ power spectrum density, following
\citet{planck2013-p03}, \citet{planck2014-a03}, and \citet{planck2016-l02},
given in Eq.~\eqref{eq:psd}.

\begin{figure}
	\includegraphics[width=\columnwidth]{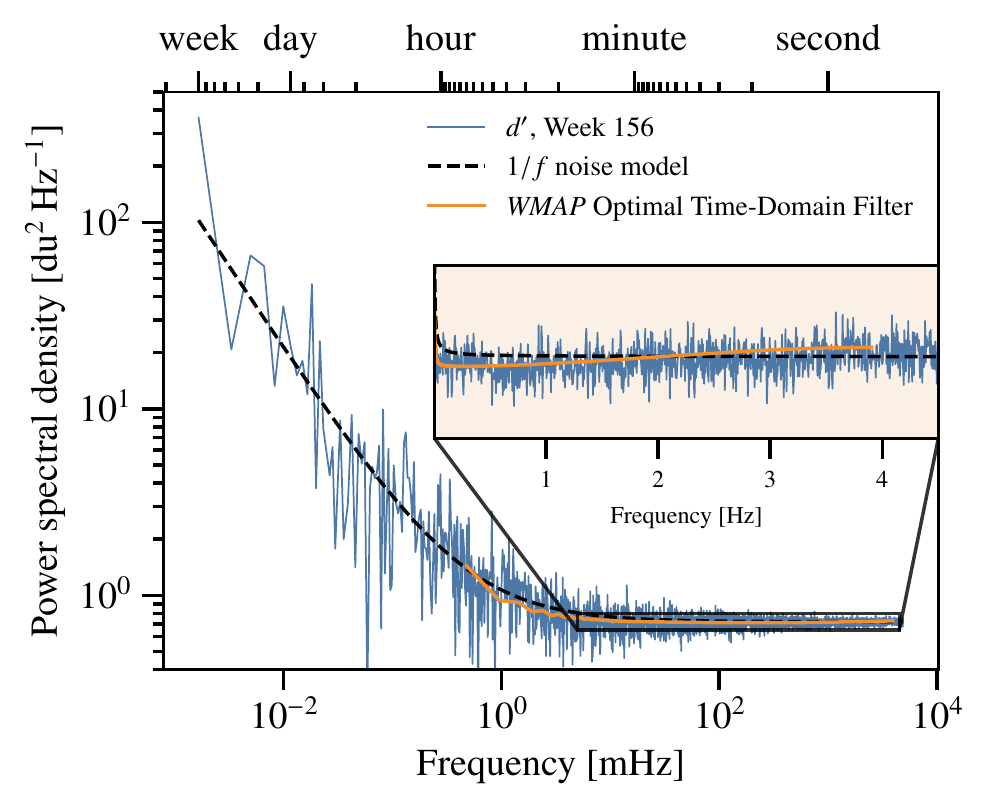}
	\caption{Power spectrum for \Q123's week 156, with \WMAP{}'s optimal time-domain filter plotted above. The inset highlights the high frequency region and the need for a non-flat noise model, as originally implemented by the \WMAP\ team.}
	\label{fig:pid_psd}
\end{figure}

Figure~\ref{fig:pid_psd} shows a comparison of these two noise models in power
spectrum domain, compared with one week of \Q123 data. Here we see two main
differences. First, the longer stationarity period of one week adopted here
allows us to model noise correlations on much longer time scales than the one
hour period adopted by \WMAP. Second, we see in the inset that the actual
\WMAP\ noise spectrum increases at high frequencies, and this is supported by
the polynomial-based \WMAP\ approach, but not by the more constrained $1/f$
noise model. Furthermore, as described by \citet{bp06}, the current noise
sampler effectively uses the highest frequencies for estimating the white noise
level to break a strong degeneracy between $f_{\mathrm{knee}}$ and $\sigma_0$.
While this works well for LFI, it is not a good approximation for \WMAP; the
$1/f$ model overestimates the noise levels at frequencies
$f>0.1\,\mathrm{mHz}$. The recoding effort required to eliminate this bias is
easy to describe but nontrivial to implement; essentially, the noise model
needs to be augmented with a component proportional to $f$, and the noise PSD sampler
needs to fit $\sigma_0$ jointly with the other parameters, not as a special
case.

However, since the main purpose of the current machinery is to perform joint
analysis of many different experiments, one must also make sure that the new
implementation supports other important near-term datasets, and most
importantly \Planck\ HFI, which also has strongly colored noise at high
frequency. Generalization of the \commander\ framework to support arbitrary colored
noise models will therefore be addressed separately in a future publication. As
far as the current work is concerned, the main impact of this shortcoming will
be biased time-domain $\chi^2$ statistics, and slightly overestimated map level
uncertainties at intermediate frequencies.

\subsection{Sidelobe corrections}
\label{sec:sidelobes}

\begin{figure}
    \includegraphics[width=\linewidth]{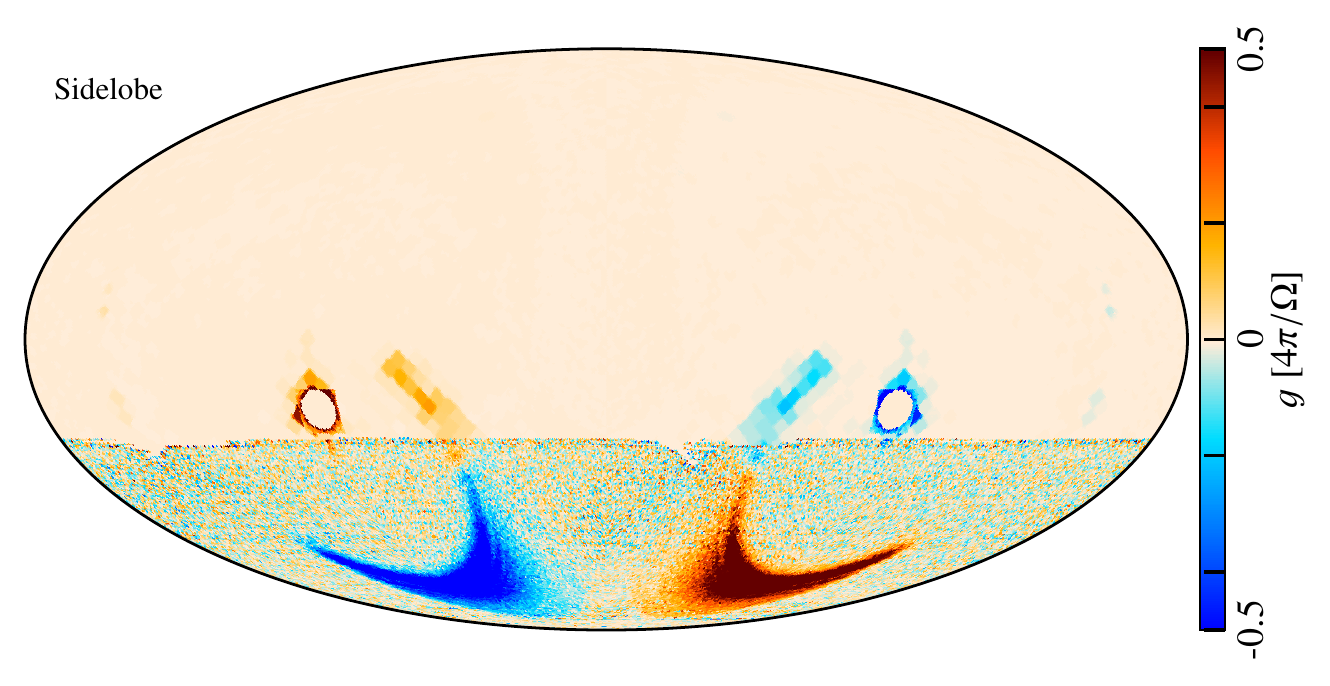}
	\caption{\WMAP\ \Q1-band sidelobe map, given in units of relative gain, $4\pi$ steradians divided by beam solid angle.}
    \label{fig:sidelobe_response}
\end{figure}

The last algorithmic step considered in this paper is far sidelobe corrections.
This formalism is identical to the \BP\ LFI analysis, as
described by \citet{bp08}, and we therefore only give a brief review of the
main points here. For a discussion of the corresponding \WMAP\ implementation,
we refer the interested reader to \citet{barnes2003}. 

The emission received by a single beam, $b(\hat{n})$, pointing toward direction
$\hat{n}=(\vartheta,\varphi)$ with a rotation angle $\psi$ may be written as
the following convolution,
\begin{equation}
  \cvar \equiv \int_{4\pi} s(\hat{n})
  b\big(\hat{n}'(\vartheta,\varphi)-\hat{n},\psi\big)\, d\Omega_{\hat{n}},
  \label{eq:convolution}
\end{equation}
where $s$ denotes the unpolarized sky signal. As shown by \citet{wandelt2001}
and \citet{prezeau2010}, this expression may be efficiently computed in
harmonic space through the use of fast recurrence relations for the Wigner
$d$-symbols, or, as shown by \citet{bp08}, in terms of spin-weighted spherical
harmonics. The main advantage of the latter is the possibility of using highly
optimized spherical harmonics libraries for the computationally expensive
parts.

In general, $b$ is a full-sky function with a harmonic bandwidth defined by the
main beam. However, this is a large and complicated object, and its structure
is typically not well characterized outside the main beam. As a result, it is common
practice to divide the full beam into (at least) two components, namely a main
beam and far sidelobes. The former is treated at full angular resolution, but
limited spatially to just a few degrees around the central axis, while the
latter is approximated with a lower resolution grid, but with full $4\pi$
coverage (except for the main beam region, which is nulled).

\begin{figure*}
	\centering
	  \includegraphics[width=0.99\linewidth]{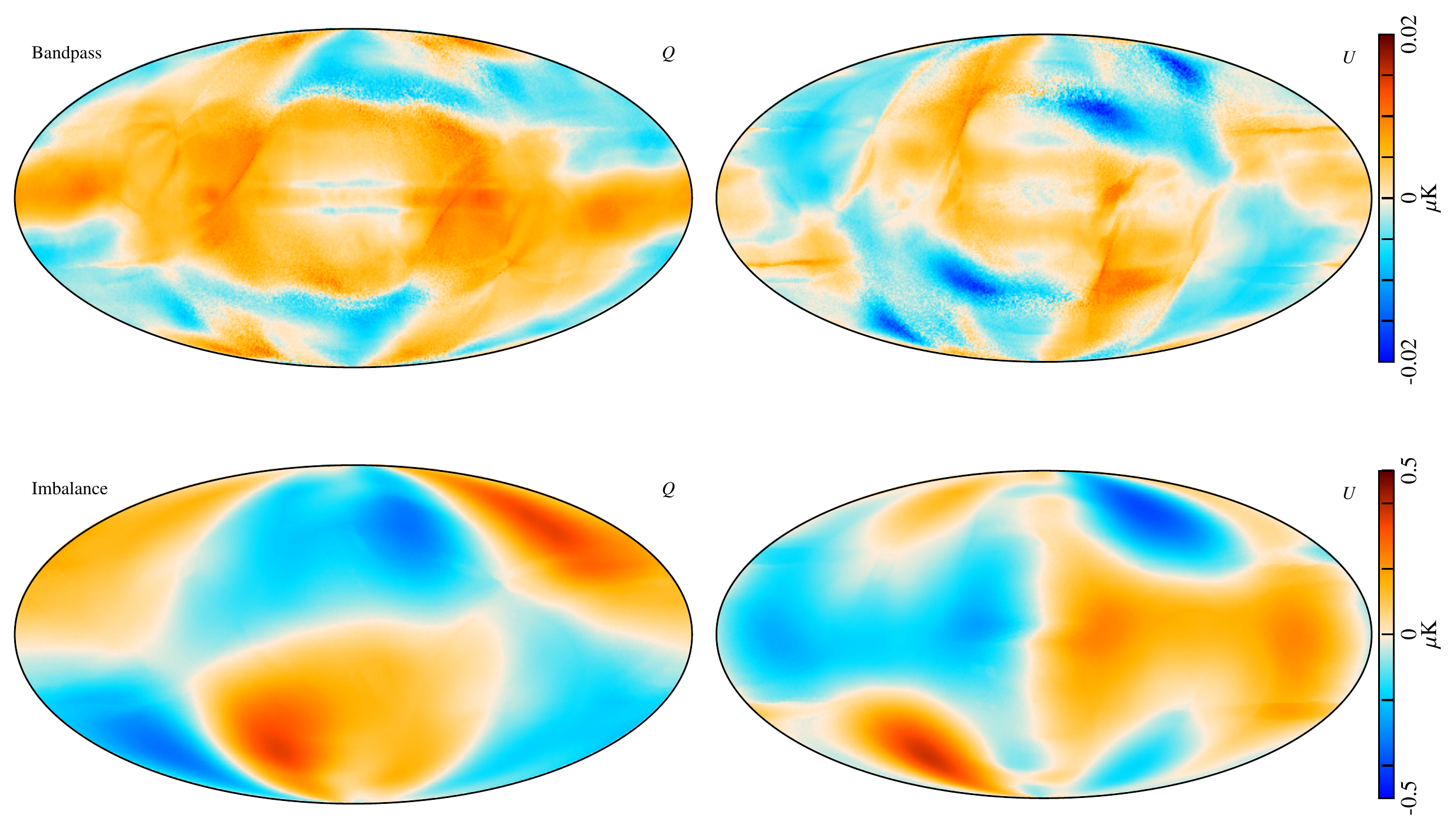}
	\caption{Estimates of \WMAP{} \Q1-band polarized sidelobe pickup. Note the differing dynamic ranges for each panel.
	(\textit{Top:}) \commander{} estimate of the polarized sidelobe pickup, taking into account only the bandpass differences between each channel.
	(\textit{Bottom:}) \commander{} estimate of the polarized sidelobe pickup, taking into account only the transmission imbalance parameters.}
	\label{fig:P_sidelobes}

%\begin{figure*}[t]
  \centering
	\includegraphics[width=0.99\linewidth]{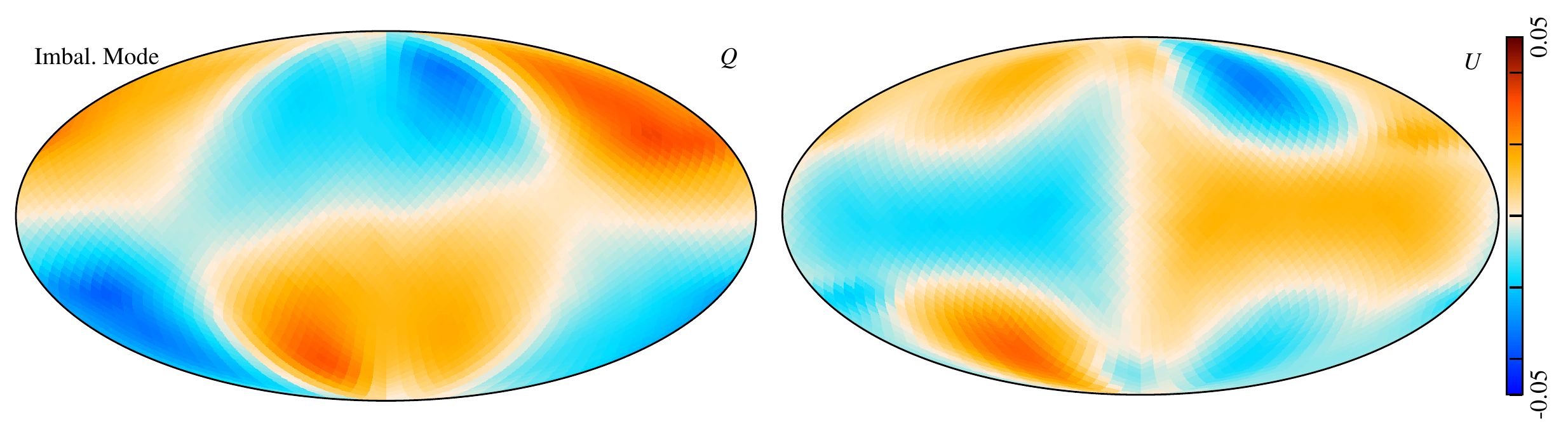}
	\caption{Official \WMAP\ \Q-band imbalance template. Note that units are arbitrary, as this is an eigenmode to be downweighted in the low-resolution likelihood analysis.}
	\label{fig:imbal_templates}
%\end{figure*}
\end{figure*}

Figure~\ref{fig:sidelobe_response} shows the far sidelobe response of the
\WMAP\ \Q1 radiometer, as estimated by \citet{barnes2003}. In this plot,
positive and negative pixels correspond to the A and B sides, respectively. The
high resolution regions are measured directly from data, using for instance
Moon observations, while the low resolution regions are estimated by model ray
tracing and laboratory measurements. The circular holes correspond to the main
beam cutouts.

Following \citet{barnes2003}, we partition this map into A and B sides
according to positive and negative pixels. 
To construct the actual sidelobe correction for a given radiometer, we
then evaluate Eq.~\eqref{eq:convolution} separately for the A and B
side radiometers, taking care to evaluate $s$ separately for each
detector. That is, the mixing matrices in Eq.~\eqref{eq:mixmat} are
integrated with respect to the bandpass of each individual radiometer. This
accounts for possible intensity-to-polarization leakage arising from bandpass
differences, which is relevant for the polarized sidelobe
corrections. 

The issue of spurious polarized signal arising from sidelobes and bandpass differences is treated in \citet{barnes2003}. Given a sidelobe model for two detectors with orthogonal polarization, $\boldsymbol s^\mathrm{fsl}_1$ and $\boldsymbol s^\mathrm{fsl}_2$, 
\begin{equation}
	\boldsymbol s_{j}^\mathrm{fsl}=\tens B_j^\mathrm{fsl}
	\mathsf M_j\,\boldsymbol a,
\end{equation}
the polarized contribution to the sidelobe signal is
\begin{equation}
	\begin{split}
	s^\mathrm{fsl,pol}_t
		&=[s_{1\A,t}^\mathrm{fsl}-s_{1\B,t}^\mathrm{fsl}]
		-[s_{2\A,t}^\mathrm{fsl}-s_{2\B,t}^\mathrm{fsl}]
	\end{split}
\end{equation}
Given an unpolarized sky, an ideal differential radiometer with identical
bandpasses and identical horn transmissions would yield no polarized far
sidelobe pickup. \citet{barnes2003} explicitly models the effect of bandpass
mismatch on the spurious polarization signal, and constrains its amplitude to be
$\lesssim0.4\,\mathrm{\mu K}$ for one year of \Q-band data. An additional
effect of transmission imbalance can induce a spurious polarization signal;
\begin{equation}
	\begin{split}
		s^\mathrm{fsl,pol}_t&=
		[(1+x_\mathrm{im,1})s^\mathrm{fsl}_{1\A,t}-(1-x_\mathrm{im,1})s^\mathrm{fsl}_{1\B,t}]
		\\
		&\,-[(1+x_\mathrm{im,2})s^\mathrm{fsl}_{2\A,t}-(1-x_\mathrm{im,2})s^\mathrm{fsl}_{2\B,t}]
		\\
		&=[s_{1\A,t}^\mathrm{fsl}-s_{1\B,t}^\mathrm{fsl}]
		-[s_{2\A,t}^\mathrm{fsl}-s_{2\B,t}^\mathrm{fsl}]
		\\
		&\,+x_\mathrm{im,1}[s_{1\A,t}^\mathrm{fsl}+s_{1\B,t}^\mathrm{fsl}]
		-x_\mathrm{im,2}[s_{2\A,t}^\mathrm{fsl}+s_{2\B,t}^\mathrm{fsl}]
	\end{split}
\end{equation}
This spurious polarization effect persists when there is no radiometer bandpass
mismatch, and in this case induces a polarized signal
$2(x_\mathrm{im,1}-x_\mathrm{im,2})(s_{\A,t}^\mathrm{fsl}+s_{\B,t}^\mathrm{fsl})$. In \citet{bennett2012},
\Q11 and \Q12 have transmission imbalance parameters of 
\begin{align}
	x_\mathrm{im,1}^\mathit{WMAP}&=-0.00013\pm0.00046
	\\
	x_\mathrm{im,2}^\mathit{WMAP}&=\phantom{-}0.00414\pm0.00025,
\end{align}
whereas our analysis has values of 
\begin{align}
	x_\mathrm{im,1}^\mathtt{Comm}&=0.00215\pm0.00026
	\\
	x_\mathrm{im,2}^\mathtt{Comm}&=0.00552\pm0.00025.
\end{align}
The difference  between the imbalance parameters
${x_\mathrm{im,1}-x_\mathrm{im,2}}$ is consistent between  both treatments,
suggesting that the imbalance-induced sidelobe polarization signal is also
present in the \textit{WMAP} timestreams.  The effect of transmission imbalance
on sidelobes is not mentioned in \citet{barnes2003}. This accounts for the main
difference between this work's polarized sidelobe estimate and that of
\citet{barnes2003}.

Figure~\ref{fig:P_sidelobes} shows the \textit Q1-band
temperature-to-polarization leakage sidelobe signal separately for bandpass
mismatch (top row) and transmission imbalance (bottom row). Here we see that
the transmission imbalance contribution is about one order of magnitude larger
than the bandpass mismatch contribution, and it has a very distinct morphology.
This morphology is very similar to the imbalance modes described in
Sect.~3.5.1 of \citet{jarosik2007}, which we reproduce in
Fig.~\ref{fig:imbal_templates}.

The similarity between the imbalance mode and the polarized sidelobe signal is
worth exploring in further detail. The two signals come from different effects.
The imbalance template was generated by evaluating the mapmaking procedure
assuming a 10\,\% increase and a 10\,\% decrease in the two imbalance
parameters, and is thus generated by small deviations from the true underlying
imbalance parameters. Conversely, the polarized sidelobe signal persists even
if the imbalance parameters are estimated, as long as $x_1\neq x_2$.

This can be partially explained by the total convolution formalism described
in \citet{bp08}. Assuming that we are convolving a beam with spherical harmonic
representation $b_{\ell m_b}$ with a sky signal $s_{\ell m_s}$ dominated by the
$\ell=1$ mode, i.e., the Solar dipole, the signal as a function of beam
orientation is given by 
\begin{equation}
	c(\vartheta,\varphi,\psi)=\sqrt{\frac{4\pi}{2\ell+1}}\sum_{m_s,m_b}
	s_{\ell m_s}b_{\ell\,-m_b}\cdot_{-m_b}Y_{\ell m_s}(\vartheta,\varphi)\e^{im_b\psi}.
\end{equation}
This means that the sidelobe pickup will mainly be determined by the $\ell=1$
modes of the beam, hence giving a large contribution that is modulated by the
boresight angle $\psi$. Conversely, incorrect imbalance templates for a pencil
beam will induce extra signal pickup from the Solar dipole that depends only on
the orientation of the spacecraft. Essentially, the signals look so similar
because they are both dominated by observing the Solar dipole with similar
orientations. At the same time, this explains the small differences between the two effects. 
The two signals are masked in the timestream at different times,
since the processing mask depends on the pointing of the main beam. This
explains why the signal is not straight along the prime meridian in the
sidelobe imbalance signal, and why the notches above and below the Galactic
anticenter in $U$ are of opposite sign between the two maps.

The main differences between the \commander\ and
the \WMAP\ sidelobe modelling presented by \citet{barnes2003}, are
the following:
\begin{enumerate}
	\item \citet{barnes2003} used a smaller transition radius between
		main beam and sidelobe of $2\fdg2$ than the final
                9-year \WMAP\ and the current analysis, both of which use
		$5\fdg0$ \citep{hill2009,bennett2012}. This causes a significant
		reduction of pickup that in later releases is included in the
		main beam. 
	\item \citet{barnes2003} generated sidelobe templates using the
		first-year scan pattern convolved with the first-year estimate
		of the sky map, whereas our analysis uses the full nine-year
		scan strategy convolved with the parametric sky model
		$\boldsymbol s^\mathrm{sky}(\nu,\boldsymbol
		a,\boldsymbol\beta)$, which has been  fit to the
		\WMAP9+Haslam+LFI+\Planck\ 353/857\,GHz sky maps. 
	\item \citet{barnes2003} explicitly took into account the polarized
		sidelobe pickup from polarized sky signal using the full
		antenna gain $G(\vec n)$ and the polarization direction
		perpendicular to the sky, $\vec P(\vec n)$. Only the gain
		amplitude is reported on
		LAMBDA,\footnote{\url{https://lambda.gsfc.nasa.gov/product/map/dr5/farsidelobe_get.cfm}}
		so we are unable to account for the intrinsically polarized pickup of the
		sidelobes. 
        \item Temperature-to-polarization leakage from transmission
          imbalance is included in the \commander\ model, but is
          not mentioned by \citet{barnes2003}.
\end{enumerate}

From the individual radiometer sidelobe corrections, we form an effective
correction per DA in the time domain, $s_{t,j}^\mathrm{fsl}$, weighting each
timestream by the respective transmission imbalance factors, as outlined in
Eq.~\eqref{eq:data_tod}. This function is then subtracted from the calibrated
TOD residual in Eq.~\eqref{eq:map_res} prior to mapmaking. Alternatively, the
entire residual may be replaced by $s_{t,j}^\mathrm{fsl}$, in which case the
resulting map will be an image of the net sidelobe correction in the map
domain. Both variations will be considered in the next section.

As a consistency check, we have developed an alternate \texttt{python} pipeline
for simulating sidelobe timestreams and
mapmaking.\footnote{\url{https://github.com/Cosmoglobe/Commander/blob/wmap/commander3/todscripts/wmap/sl_conv.py}}
We use the \texttt{ducc}\footnote{\url{https://gitlab.mpcdf.mpg.de/mtr/ducc}}
implementation of \texttt{totalconvolver} \citep{wandelt2001,prezeau2010} to
simulate the \WMAP\ sidelobe contribution from a pure Solar dipole reported by
\citet{jarosik2010}, with amplitude $3355\,\mathrm{\mu K}$ and direction
${(l,b)=(263\fdg99,48\fdg26)}$ and the two horns' orientations.  We use the
transmission imbalance parameters reported in \citet{bennett2012} to simulate
observed timestreams, then bin the maps at $N_\mathrm{side}=16$ and solve for
the output map exactly using the \texttt{scipy} sparse linear algebra package
\citep{scipy2020}. This \texttt{python} implementation of the mapmaker and
simulated timestreams reproduces a low resolution version of maps obtained
using \commander's iterative solver.

\subsection{Differences between the \cosmoglobe\ and \WMAP\ pipelines}
\label{sec:differences}

% Should this section go closer to the conclusions?

Although our goal is not a reproduction of the \WMAP{} mapmaking pipeline, we
have in general attempted to follow the \WMAP{} team's approach.  There are
some algorithmic differences that can result in different frequency maps
derived from the same time-ordered data. The differences we are aware of include
the following:
\begin{itemize}
\item \emph{Calibration}: The \WMAP\ team developed a physical
  model for the gain of the instrument as a function of housekeeping
  parameters. These parameters were fit to the hourly gain estimates
  using the CMB dipole as a calibration source. The 9-year best-fit
  parameters are given in Appendix~A of \citet{wmapexsupp}. In
  contrast, the \commander\ framework calibrates the time-ordered data on a
  weekly cadence by comparing directly to the expected sky amplitude,
  and most importantly to the orbital and Solar CMB dipoles; for full
  details, see \citet{bp07}. The \commander\ approach makes fewer
  assumptions regarding the physical origin of gain fluctuations, but
  stronger assumptions regarding their stability in time.
\item \emph{Transmission imbalance}: As discussed by
  \citet{jarosik2003,jarosik2007}, the \WMAP\ team derived
  transmission imbalance parameters using ten precession periods at a time, and
  estimated a corresponding uncertainty through the variation during
  the mission. This quantity is also dependent on the treatment of
  low-frequency noise, so the original algorithm solved for  a cubic polynomial in each period
  while fitting for the transmission imbalance coefficients. The
  resulting transmission imbalance templates were projected out from
  low-resolution noise covariance matrices to account for their effect
  on cosmological parameters.  In contrast, the current analysis assumes a constant
  imbalance parameter for the entire mission, but allows this to vary
  in each Gibbs iteration, and thereby marginalizes over this parameter
  directly in the data model, fully analogous to the gain
  \citep{bp07}. We apply no explicit postproduction corrections to any data
  product for transmission imbalance.
\item \emph{Baseline evaluation}: The raw \WMAP\ data have a large
  offset from zero. This is treated explicitly as a time-varying
  constant within each stationary period in the official pipeline, but
  as an offset in the correlated noise component in the \commander\
  framework. In the picture where baseline and correlated noise are
  treated as separate terms, there is a correlation between gain,
  baseline, and $1/f$ noise, as discussed in \citet{hinshaw2003a}. To
  address this, the \WMAP{} team applied a whitening filter using an
  estimate of the $1/f$ noise spectrum, and iteratively solved for the
  baseline. In contrast, we fit a linear baseline per weeklong scan in 
  the \commander\ pipeline, and residual baseline fluctuations are absorbed
  by the $\boldsymbol n^\mathrm{corr}$ sampling \citep{bp06}.
\item \emph{Solar dipole}: While the \WMAP\ team subtracted an estimate of the
  Solar dipole from the time-ordered data before mapmaking, resulting
  in dipole-free frequency maps \citep{hinshaw2003a}, \commander\ retains it
  for calibration and component separation purposes, following
  \Planck\ DR4 \citep{npipe,bp07,bp13}.
\item \emph{Mapmaking}: The \WMAP\ team accounted for correlated noise weighting
  by estimating a two-point correlation function in time-domain, and
  used this to pre-whiten the TOD prior to mapmaking. In contrast,
  we assume a $1/f$ noise model, and sample correlated noise
  explicitly as a stochastic field in time-domain \citep{bp06}. We
  note again that this noise PSD model will be generalized in the
  future to fully capture the temporal behaviour of the \WMAP\ noise,
  as discussed in Sec.~\ref{sec:psd}.
\item \emph{Bandpass corrections}: The \WMAP\ team suppressed
  temperature-to-polarization leakage from bandpass differences
  between radiometers by solving for a spurious map, $S$, per
  radiometer. In contrast, we use a parametric
  foreground model to subtract these bandpass effects in time-domain
  \citep{bp09}. The main advantage of the latter is that it requires fewer free
  parameters, and therefore results in a lower degree of white noise, while the
  main disadvantage is a stronger sensitivity to the foreground
  model. We find that the two approaches perform similarly. 
\end{itemize}

\section{Results}
\label{sec:Results}

\begin{table}
	\caption{Computational resources for end-to-end \wmap\ \Q1-band processing. The values are given in total CPU-hours averaged over 100 samples. Wall time is obtained by dividing by $n_\mathrm{cores}=64$. Note that auxiliary maps are computed every tenth run, contributing to a higher average runtime.}
    \label{table:timing}      % is used to refer this table in the text
    \centering                                      % used for centering table
    \begin{tabular}{l r r}
    \hline\hline                        % inserts double horizontal lines
    \textsc{Item} & \textsc{Cost} & \textsc{Percentage}\\
    \hline                                             %inserts single line
    \textit{Data volume}
    \\
    \quad Uncompressed volume & 76 GB &
    \\\vspace*{1mm}
    \quad Compressed volume & 14 GB &
    \\
    \textit{Processing time (cost per run)}
    \\
    \quad TOD initialization/IO time & 0.3\,h &
    \\\vspace*{1mm}
    \quad Other initialization & 2.2\,h &
    %\\
    %\quad \textbf{Total initialization} & 165\,s
    \\
    \textit{Processing time (cost per sample)} & 
    \\
    \quad Data decompression & 0.8\,h & 3.5\,\%
    \\
    \quad TOD projection ($\textsf P$ operation) & 1.1\,h & 4.9\,\%
    \\
    \quad Sidelobe precomputation & 0.2\,h & 0.4\,\%
    \\
    \quad Sidelobe interpolation & 6.4\,h & 29.4\,\%
    \\
    \quad Orbital dipole & 1.4\,h & 6.4\,\%
    \\
    \quad Gain sampling & 1.0\,h & 4.6\,\%
    \\
    \quad Transmission imbalance sampling & 0.3\,h & 1.2\,\%
    \\
    \quad Correlated noise sampling & 1.8\,h & 8.2\,\%
    \\
    \quad Correlated noise PSD sampling & 0.4\,h & 2.0\,\%
    \\
    \quad Map making & 6.0\,h &  27.3\,\%
%    \\
%    \quad Time per BiCG-STAB iteration & 30\,s
%    \\
%    \quad Loss due to poor load-balancing
    \\\vspace*{1mm}
    \quad Sum of other TOD steps & 2.1\,h & 9.8\,\%
    \\
%    \quad TOD processing per sample 
%    \\
    \quad \textbf{Total cost per sample} & 21.8\,h & 100.0\,\%
    \\
    \hline
\end{tabular}
\end{table}

We are now ready to present the results obtained by applying the methods
outlined in Sects.~\ref{sec:bp} and \ref{sec:algorithms} to the \WMAP\ \Q1-band
data. We have made all data, code, and parameter files required to perform this
analysis publicly available in the \commander\ source
code.\footnote{\url{https://github.com/Cosmoglobe/Commander/tree/wmap}}

We emphasize that the results presented below are a simplified version of the
more complete Gibbs sampling problem. Similar to the \WMAP9 analysis, we
calibrate the raw data to a fixed signal and adjust the gain parameters and
noise model while correcting for known instrumental effects. Our analysis
differs from the \WMAP9 processing in that we calibrate to the total sky model
rather than just the orbital dipole. The analysis choice to enforce a single
sky model across all sky maps is the primary advantage of the \commander\
method, by design reducing degeneracies due to observing strategy. A complete
\commander\ analysis, jointly analyzing the time-ordered data of both \Planck\
LFI and \WMAP\ while fitting for the sky parameters, will fully leverage the
power of this method.

\subsection{Computational resources}

\begin{figure}
  \centering
	\includegraphics[width=\linewidth]{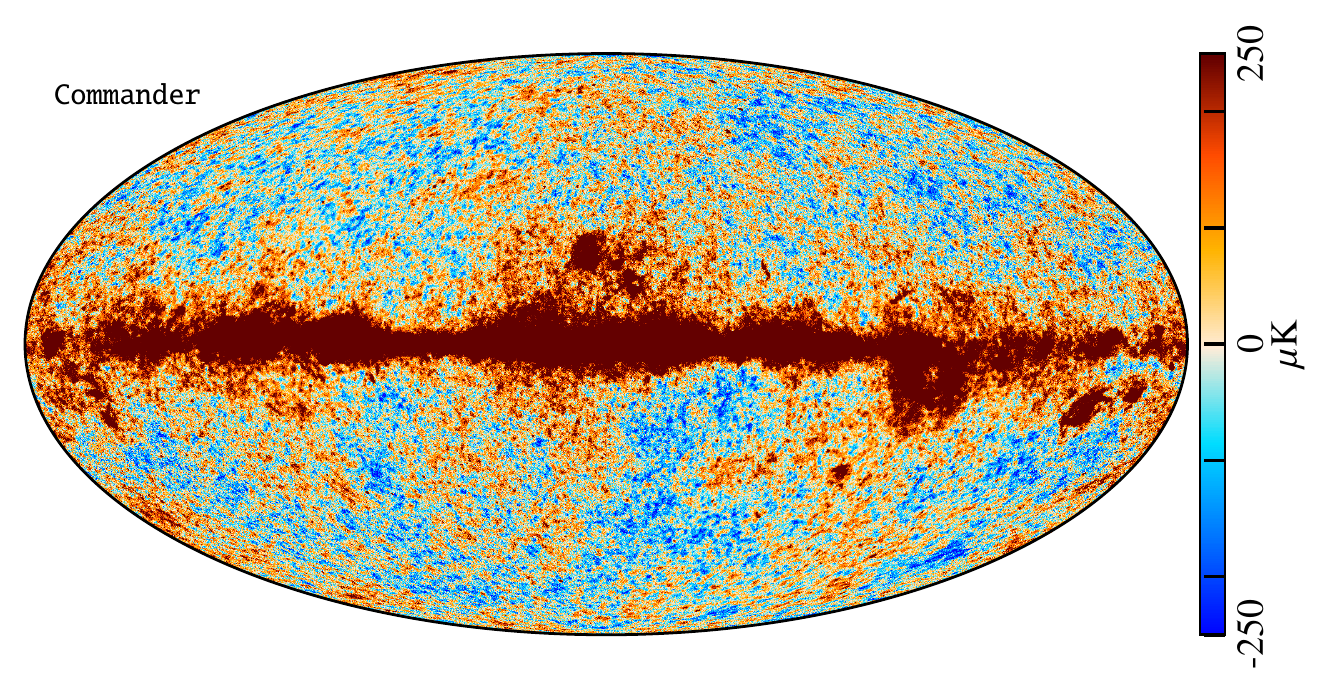}\\
	\includegraphics[width=\linewidth]{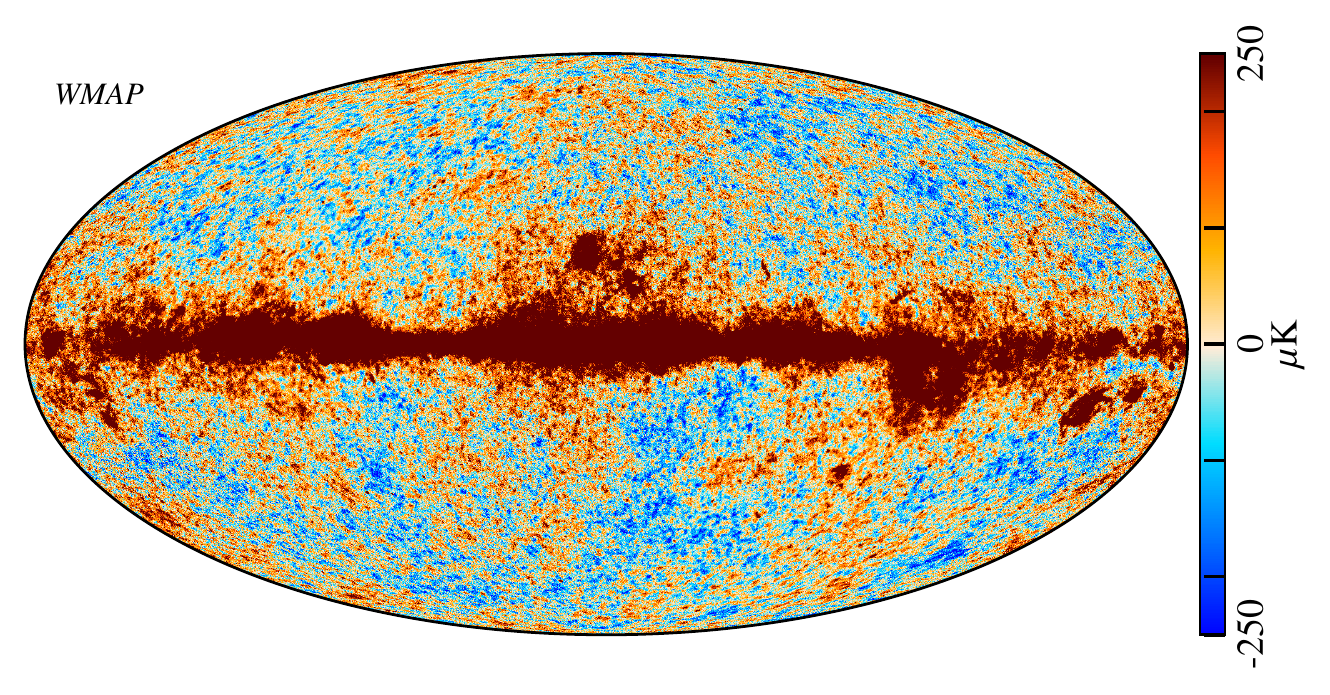}\\
	\includegraphics[width=\linewidth]{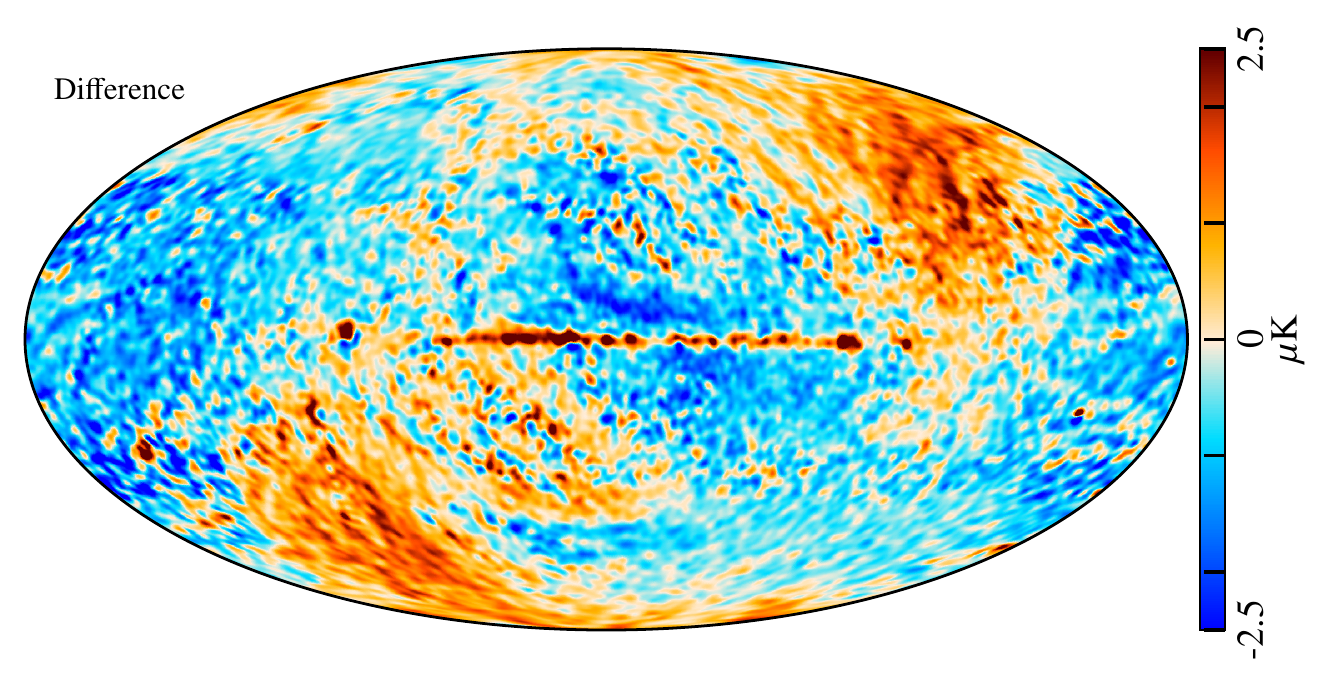}\\
	\includegraphics[width=\linewidth]{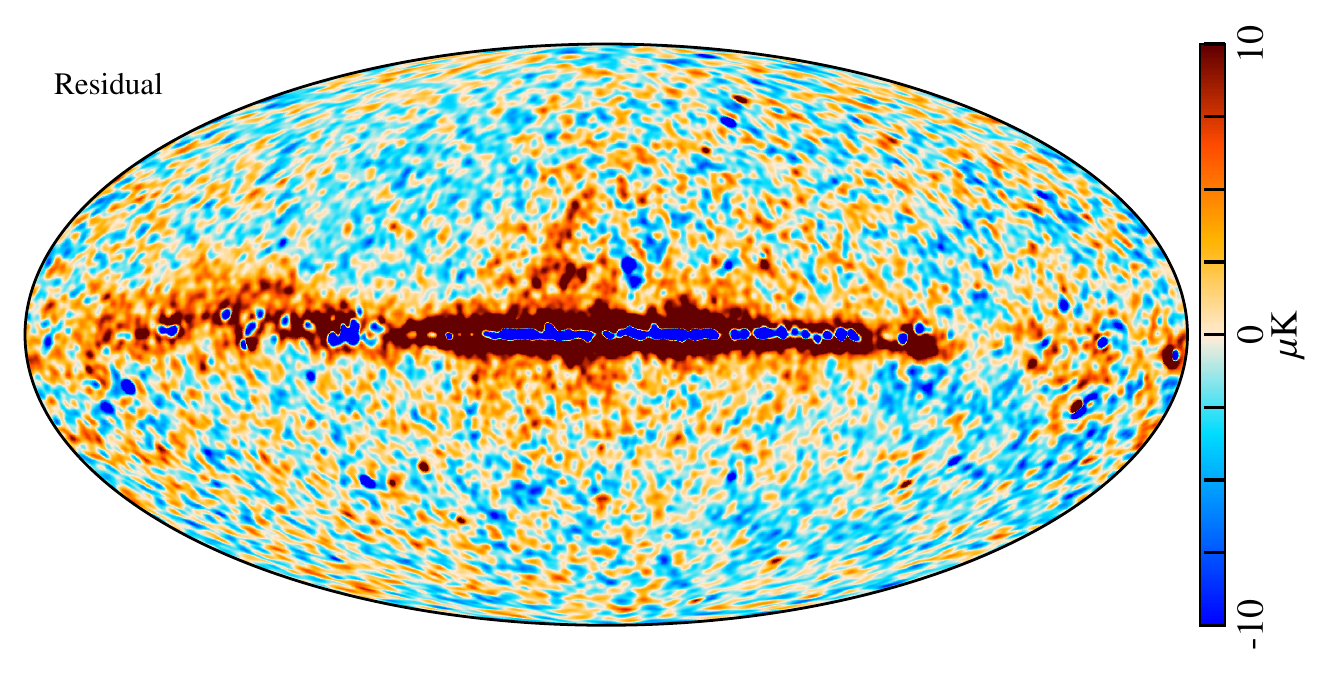}\\
	\includegraphics[width=\linewidth]{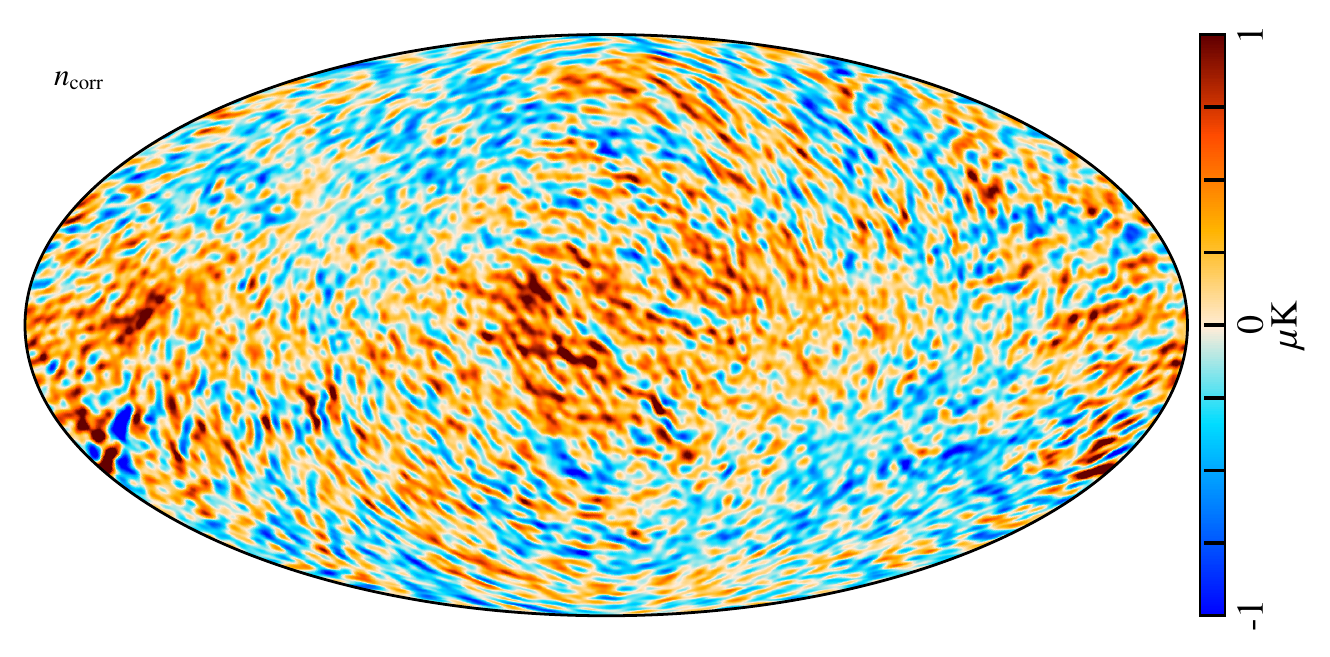}        
	\caption{(\textit{Top panel:}) \commander\ \Q1-band temperature
          map. (\textit{Second panel:}) Corresponding official
          \WMAP\ sky map. (\textit{Third panel:}) Straight difference
          between \commander\ and \WMAP. (\textit{Fourth panel:}) Single
          \commander\ TOD residual sample. (\textit{Bottom panel:})
          Single \commander\ correlated noise sample.
            }
	\label{fig:skymaps}
\end{figure}

\begin{figure}
	%\begin{center}
	%\includegraphics[width=\linewidth]{diff_Q_noquad.pdf}
	%\end{center}
	%\hbox{\hspace{-0.0em}\includegraphics[clip, trim=139mm 224.1mm 13mm 18mm,
	%  width=0.9\linewidth]{ba_f03.pdf}\hspace{2.3em}}\\
	\includegraphics[width=0.94\linewidth]{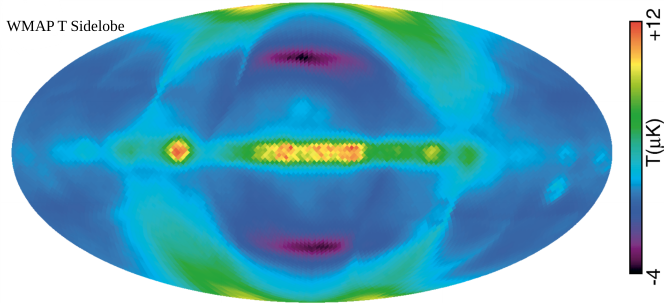}
	\begin{center}
  	\includegraphics[width=\linewidth]{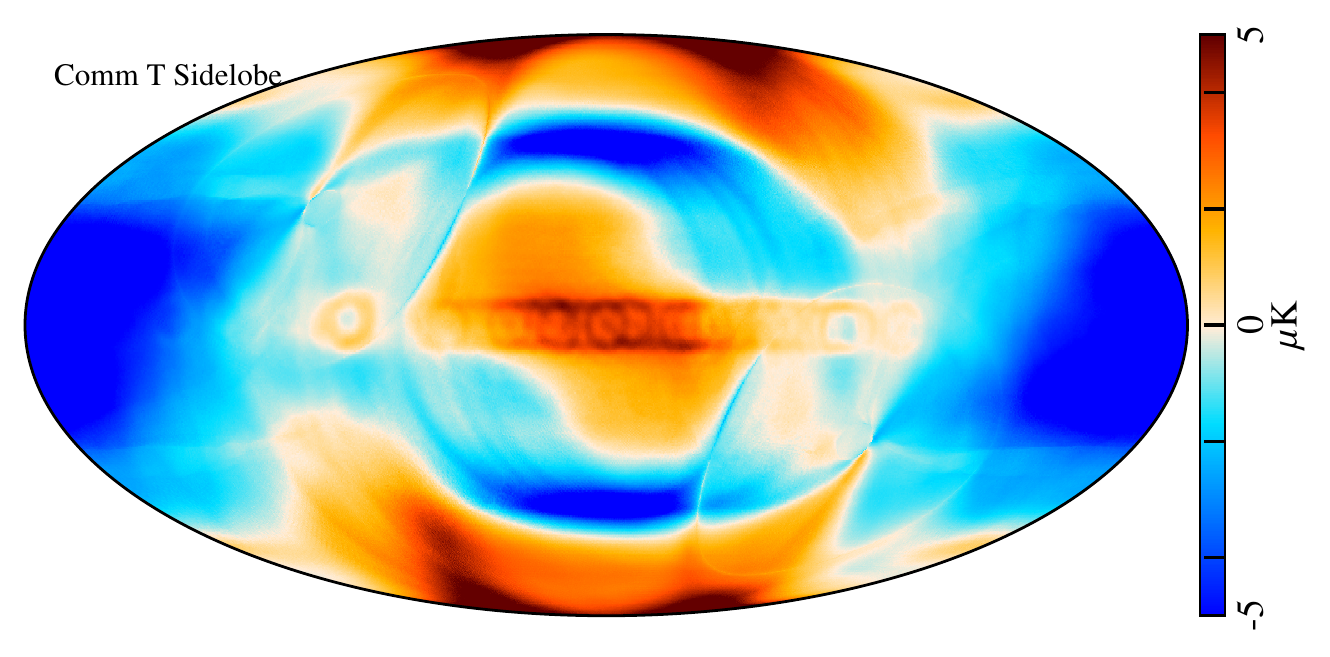}
	\end{center}
	\caption{(\textit{Top:}) 
          \WMAP\ estimate of the temperature sidelobe correction;
          reproduction from \citet{barnes2003}. (\textit{Bottom:})
          \commander\ estimate of the temperature sidelobe correction, after an overall dipole is removed.}
	\label{fig:T_sidelobes}
\end{figure}

The first main goal of the current paper is to quantify the computational
resources that will be required for a future full end-to-end \WMAP\ analysis.
Most of our efforts have been spent on implementing the main new analysis
steps, rather than optimizing for speed. Examples of known optimizations left
for future work include using optimal lengths for Fast Fourier Transform
evaluations \citep{bp03}, implementing a more effective preconditioner for
mapmaking \citep{bennett2012}, and improving load balancing for
parallelization.  With these caveats in mind, we summarize in
Table~\ref{table:timing} the computational expenses required to run the
\Q1-band analysis on a computer system with 64 AMD EPYC3~7543 2.8~GHz cores,
using the Intel Parallel Studio XE 20.0.4.912 Fortran compiler.  The processing
times listed include the integrated time including each parallel core. To
obtain the cost in wall time, the cost must be divided by the number of cores
used, in this case 64.

The total cost per full Gibbs sample is 22\,CPU\nobreakdash-hrs, which is
comparable to the \Planck\ LFI 44\,GHz channel cost of 17\,CPU\nobreakdash-hrs
\citep{bp03}. This is despite the fact that the compressed \Q-band data volume
is only 8\,\% of the 44\,GHz, and all time-domain operations are
correspondingly faster. The explanation is the more expensive differential
mapmaker; more than half of the total \WMAP\ analysis time is spent in the
mapmaking procedure, evaluating the matrix operation $\mathsf
P^T_\mathrm{am}\mathsf N^{-1}\mathsf P$. For comparison, map binning for the
LFI 44\,GHz channel accounts for less than 3\,\% of its total run time. Thus,
the differential mapmaking procedure is the main additional step in the \WMAP\
analysis, and all other processing steps scale with $\mathcal O(N\log N)$,
due to the use of FFT's in the calibration steps.  Better preconditioning is a
priority for optimizing the current code.  Based on a conservative $\mathcal
O(N\log N)$ scaling, we expect the cost for each band to be 20, 20, 30, 40, and
72 CPU\,hrs for each \K, \Ka, \Q, \V, and \W\ DA, respectively.

The memory requirement for storing the \Q1-band data is only 14\,GB, a factor
of four reduction of the uncompressed 76\,GB data. The full \WMAP\ data volume
only corresponds to 20\,\% of the LFI data volume, which can be considered an
incremental increase. Based both on memory and CPU requirements, we conclude
that a future Bayesian end-to-end \WMAP\ analysis is well within the reach of
current computer systems, both as a standalone analysis and as a joint
\WMAP--LFI analysis. 

\subsection{Temperature map quality assessment}

\begin{figure*}
  \centering
	\includegraphics[width=0.99\linewidth]{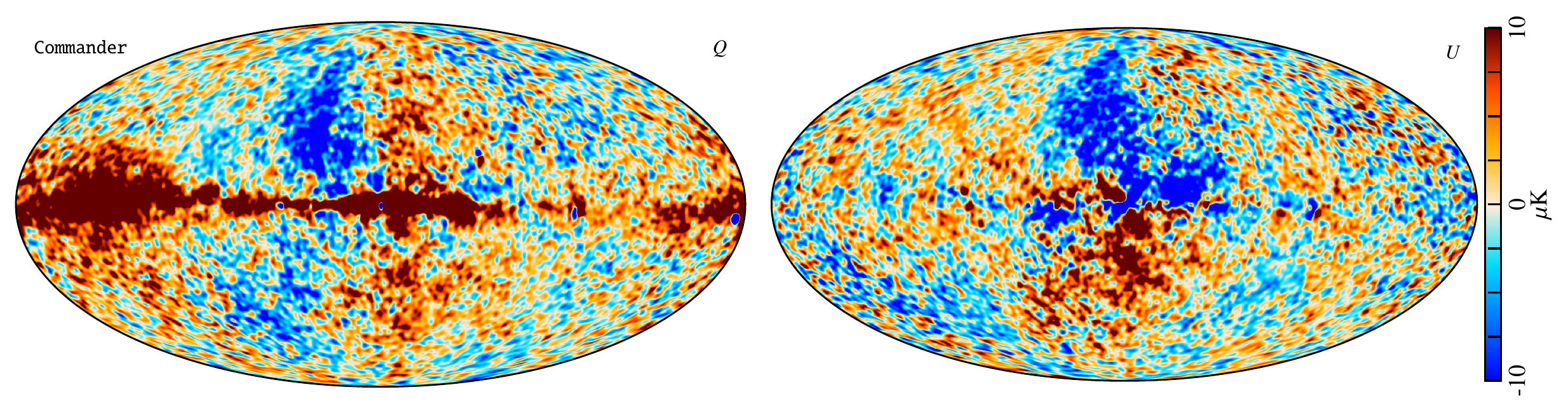}\\
	\includegraphics[width=0.99\linewidth]{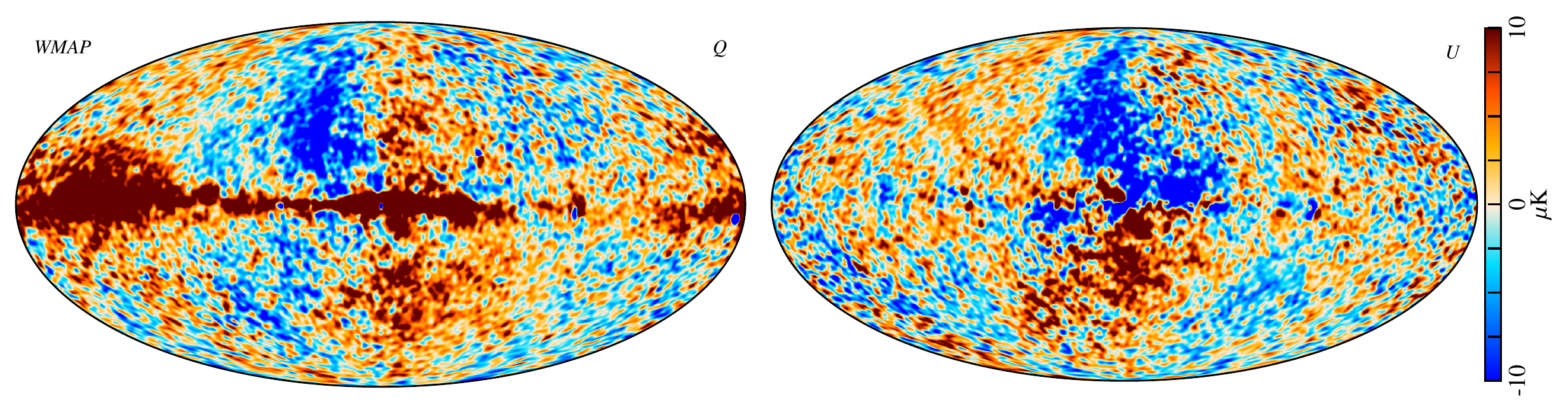}\\
	\includegraphics[width=0.99\linewidth]{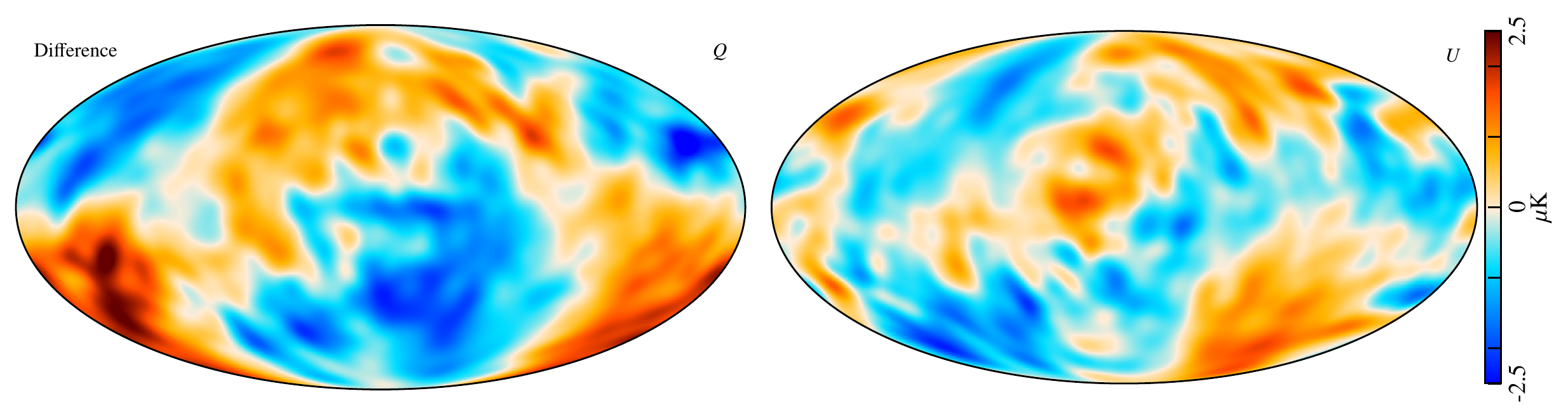}
	\caption{(\textit{Top:}) \commander{} \Q1-band polarization
          map, smoothed to $2^\circ$ FWHM. (\textit{Middle:}) Corresponding official \WMAP\ \Q1-band
	  sky map, also smoothed to $2^\circ$ FWHM. (\textit{Bottom:}) \commander{} solution minus \WMAP\ solution, smoothed to $10^\circ$ FWHM.}
	\label{fig:Pskymaps}
\end{figure*}

We now turn to the quality of the maps, starting with the temperature
component. The top panel in Fig.~\ref{fig:skymaps} shows the \commander-derived
\Q1-band sky map (after subtracting the Solar CMB dipole), the second panel
shows the corresponding official \WMAP9\ sky map \citep{bennett2012}, and the
third panel shows their difference. Overall, we see that the two maps are
consistent at the $\sim2.5\,\mathrm{\mu K}$ level with several systematic
features, of which the most salient is a quadrupole aligned with the Solar
dipole.

This signal does not appear in either the \commander\ TOD residual map (created
by binning $\d-\s^{\mathrm{tot}}-\n^{\mathrm{corr}}$ into a sky map; fourth
panel in Fig.~\ref{fig:skymaps}), or the correlated noise map (bottom panel).
Together, these two maps act as a ``trash can'' in the \commander\ processing
framework, in the sense that they highlight any signal in the raw data that
cannot be accommodated by any of the other components. The main structure in
the residual and $\n^\mathrm{corr}$ maps is correlated with the Galactic plane,
indicating an inadequate sky model.

\begin{figure*}
  \centering
	\includegraphics[width=0.99\linewidth]{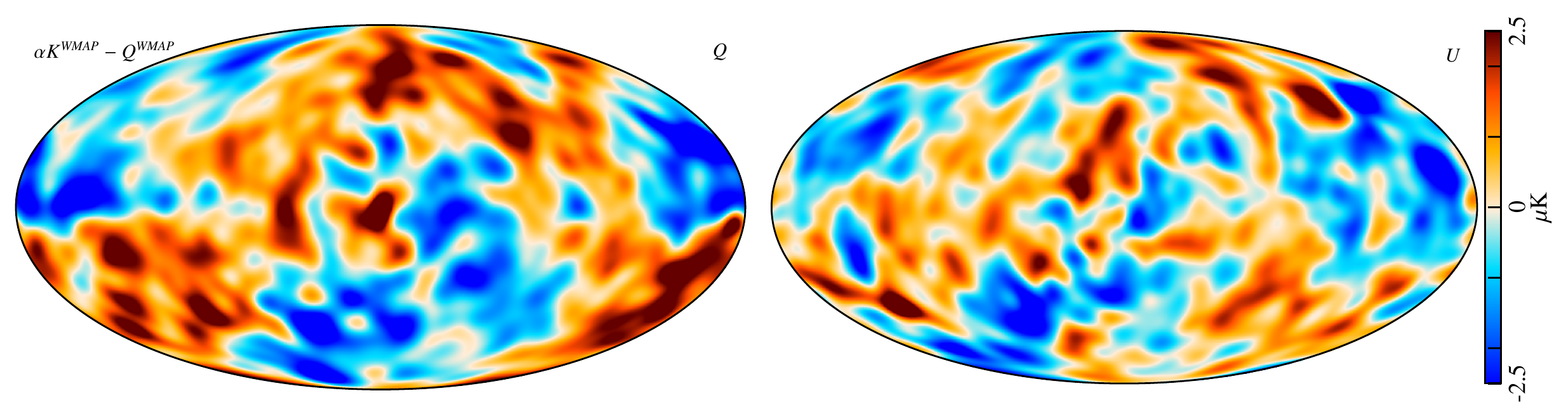}\\
	\includegraphics[width=0.99\linewidth]{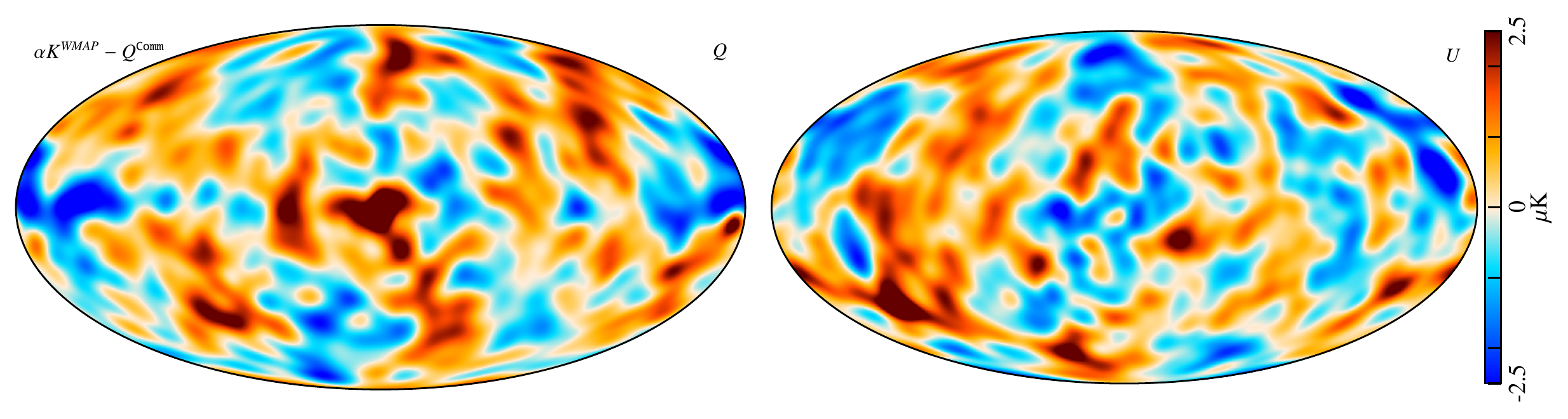}
	\caption{Difference maps between \WMAP\ \K-band and \WMAP\
          (\emph{top row}) and \commander\ (\emph{bottom row})
          \Q1-band, designed to reduce polarized synchrotron
          emission. The \K-band map has been scaled by a factor of
          0.17 to account for the different central frequencies
          of the two frequency channels. }
	\label{fig:K_minus_Q}
\end{figure*}        

\begin{figure*}
  \centering
	\includegraphics[width=0.99\linewidth]{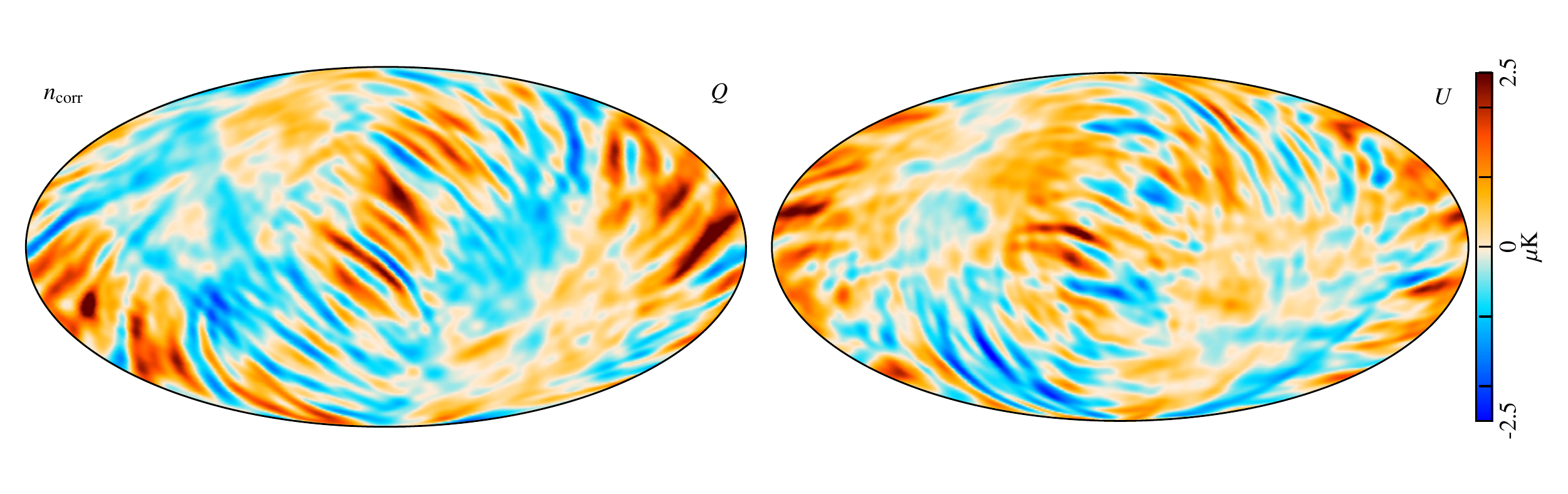}\\
	\includegraphics[width=0.99\linewidth]{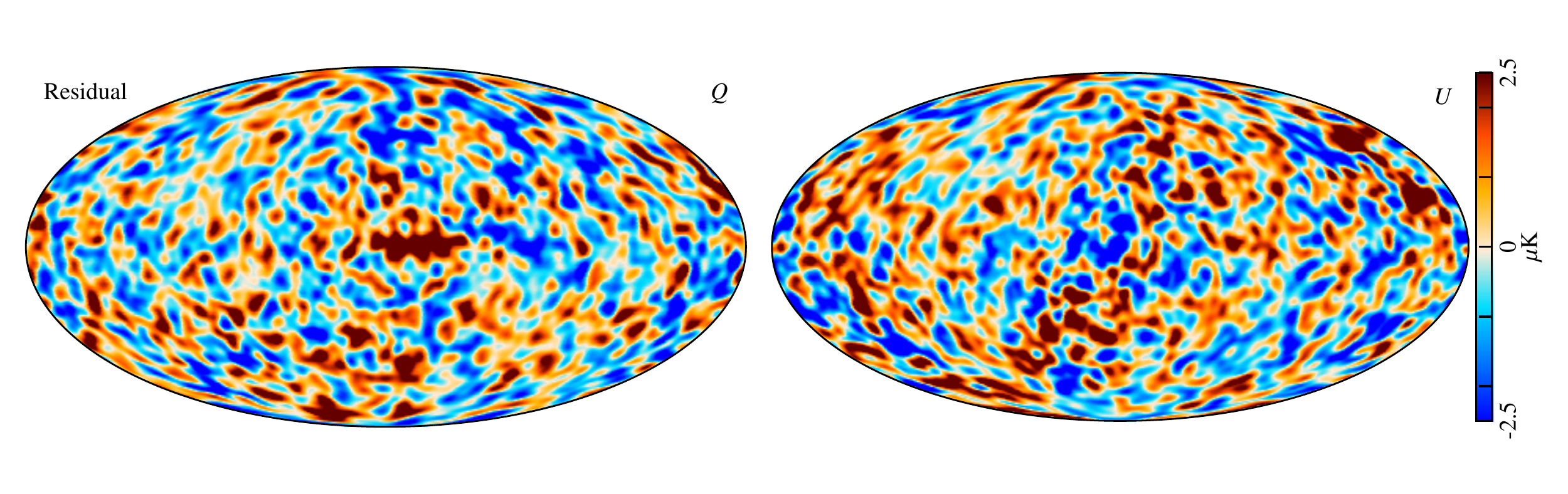}
	\caption{(\emph{Top row:}) \commander\ correlated noise sample
        for the \Q1 channel in polarization smoothed with $5\deg$ FWHM. (\emph{Bottom row:}) Corresponding TOD residual sample, also smoothed with a $5\deg$ FWHM beam. }
	\label{fig:P_gof}
\end{figure*}

We reproduce the \Q-band intensity sidelobe in Fig.~\ref{fig:T_sidelobes}, and
compare with the dipole-subtracted sidelobe prediction in the bottom panel. The
amplitude and morphology of the two computations are very similar, except for
the holes in point sources around the Galactic plane. This is due to both the
different amounts of data used in the two, one and nine years respectively, and
to the different sidelobe cutoff radii in both sidelobe maps, $2\fdg2$ and
$5\fdg0$, respectively.\footnote{ The initial version of this work had a
difference between the \WMAP\ final sky map and this reanalysis's sky map that
was morphologically identical to the \WMAP\ sidelobe as computed by
\citet{barnes2003}. We have since determined that we oriented the beam incorrectly.
We have corrected this by rotating by $135^\circ$ before convolving with the sky
model.  }

\subsection{Polarization map quality assessment}

We now turn our attention to the polarization sky maps. As in
Fig.~\ref{fig:skymaps} for temperature, the top panel of
Fig.~\ref{fig:Pskymaps} shows the \commander\ \Q1-band Stokes $Q$ and $U$ maps,
the middle panel shows the corresponding official \WMAP\ sky map, and the
bottom panel shows their difference. We see that the Galactic plane is in this
case almost perfectly consistent between the two pipelines, but there is also a
distinct large-scale pattern present at high Galactic latitudes with a
morphology similar to the signal that the \WMAP\ team identified as poorly-measured
modes in the mapmaking procedure, which can be seen in the \WMAP\
official imbalance templates shown in Fig.~\ref{fig:imbal_templates}. This structure was
identified in Sect.~3.5.1 of \citet{jarosik2007} as the coupling of the dipole
signal with small errors in the transmission imbalance parameters. However, as
discussed in Sect.~\ref{sec:sidelobes}, a major novel result from the current
analysis is that a nearly identical morphology may be reproduced
deterministically in terms of temperature-to-polarization leakage arising from
the three-way coupling between the CMB Solar dipole, transmission imbalance,
and sidelobe pickup. 

In particular, Fig.~\ref{fig:P_sidelobes} shows the polarized sidelobe
predicted by \commander, using the model described in
Sect.~\ref{sec:sidelobes}.  We note that the amplitude of the polarized
sidelobe predicted by \commander\ is an order of magnitude smaller than the
large scale feature difference feature. The amplitude difference may be caused
by the sidelobe itself or the magnitude of the transmission imbalance factors.
This implies that the polarized sidelobe as described here can at best only
partially explain the difference between the \commander\ and \WMAP9 maps.

Next, we want to understand whether the residual pattern in the bottom panel of
Fig.~\ref{fig:Pskymaps} is present in the \commander\ or \WMAP\ maps, or both.
To this aim, Fig.~\ref{fig:K_minus_Q} shows differences between the \WMAP\
\K-band channel and the \WMAP\ (top) and \commander\ (bottom) \Q1-band maps, in
which polarized synchrotron emission is greatly suppressed. In both cases, the
\K-band map has been scaled with a factor of 0.17 prior to subtraction, to
account for the different central frequencies of the two channels, equivalent
to assuming a synchrotron spectral index of $\beta_{\mathrm{s}}=-3.1$. In these
plots, we see that the sidelobe pattern is present in the official \WMAP\
\Q1-band map, while it is much weaker in the \commander\ map. This implies that
\commander\ is able to remove the imbalance modes accurately without the
need of a post-producution modification.

\begin{figure*}
  \centering
	\includegraphics[width=0.99\linewidth]{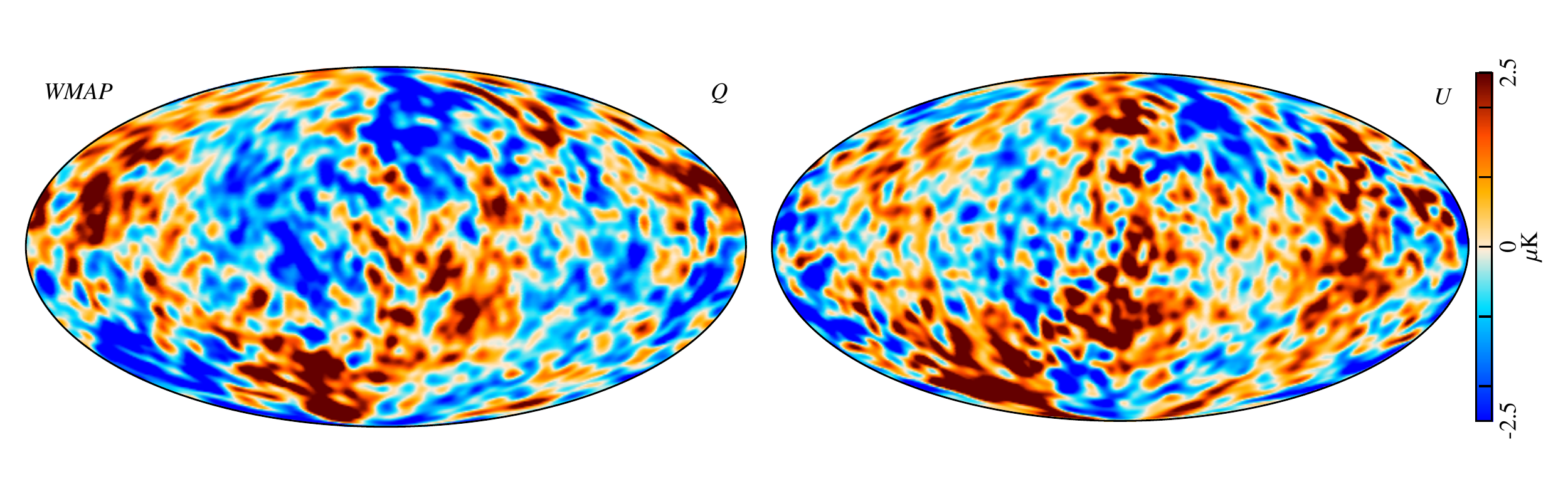}\\
	\includegraphics[width=0.99\linewidth]{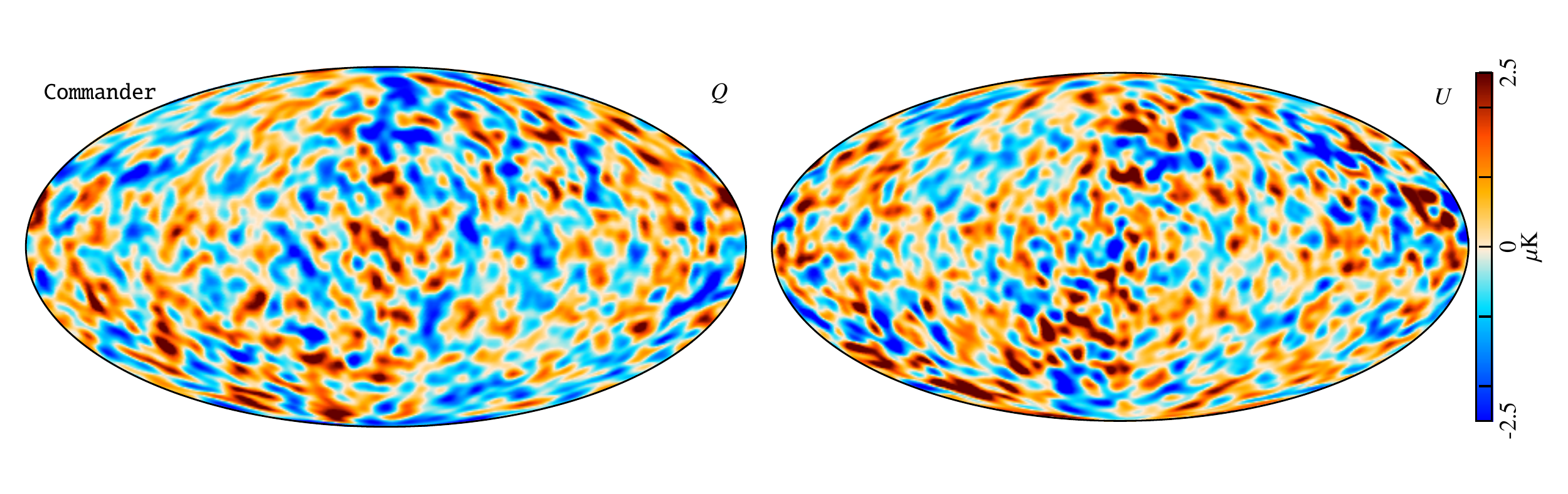}
	\caption{Half-difference maps $(\mathit{Q}1-\mathit{Q}2)/2$, smoothed to $5^\circ$. (\emph{Top row:}) \WMAP\ 
	half-difference map. (\emph{Bottom row:}) \texttt{Commander} half-difference map. }
	\label{fig:P_qdiff}
\end{figure*}

The only possible location for the imbalance modes in the \texttt{Commander}
data model is in the correlated noise, which we display a map of along with the
TOD residuals in Fig.~\ref{fig:P_gof}.  Here we see large-scale structures with
a similar amplitude to the imbalance modes in \WMAP.  The imbalance mode and
polarized sidelobe are accounted for in the correlated noise component for
\commander, resulting in broad features modulated by the scanning strategy.
Ideally, this should look like stochastic, ``stripy'' noise \citep[see,
e.g.,][for an LFI example]{bp10}, but this map is both systematic and
inconsistent with the $1/f$ correlated noise model. The reason for why this
signal does not end up in the actual frequency sky map is that we assume the
sky model to be defined by the \Planck- and LFI-based \BP\ model, which
provides the necessary leverage to extract the current signal. We could find no
compelling evidence for the imbalance modes being present in the correlated
noise map at the $2.5\,\mathrm{\mu K}$. However, preliminary runs of the other
DAs have have this signature in the correlated noise level at a higher level,
so we do not yet rule out the imbalance modes being absorbed into the
\Q1 DA's correlated noise.

As a final test of our method, we directly compare the polarized \Q1 and
\Q2-band data. The two DAs have overlapping bandpasses, with effective
frequencies of 40.72\,GHz and 40.51\,GHz, respectively \citep{bennett2012}.
Several other instrumental properties, such as the knee frequencies, gain, and
beam orientation, vary between the two DAs. Properly treated, the maps derived
from these DAs should be consistent with each other, save for instrumental
noise fluctuations. At the same time, the consistency of the \WMAP-processed
\Q\ DAs compared with the \commander-processed \Q\ DAs highlights the
differences between our two approaches. 

Figure~\ref{fig:P_qdiff} shows the polarized half-difference maps, ${(\mathit
Q1-\mathit Q2)/2}$, for the $Q$ and $U$ Stokes parameters. As expected, the
\WMAP\ half-difference maps are visually dominated by the poorly-measured imbalance modes.
As explained in \citet{jarosik2007},
this is mainly a large-scale phenomenon that is downweighted in the
low-resolution likelihood analysis. Conversely, the \commander\ half-difference
maps are consistent at the $2.5\,\mathrm{\mu K}$ level without any
post-processing. This agreement is partially due to conditioning the data on
the sky model, so that correlated noise accounts for the deviations from the
sky model. Ideally, \WMAP\ approach of downweighting imbalance modes in
the likelihood should be mathematically equivalent to the \commander\ approach
of drawing random samples of these modes to marginalize over them.

\section{Summary and conclusions}
\label{sec:Conclusions}

This paper had two main goals. The first goal was to assess whether a full
Bayesian end-to-end \BP-style analysis of the \WMAP\ data is technically
feasible, and, if so, how expensive it would be. Based on the results presented
here, we can conclude affirmatively, as the current computational cost is
44\,CPU\nobreakdash-hrs per full \Q1-band Gibbs sample. This is comparable to
the cost of the \Planck\ LFI 44\,GHz channel, and thus well within the reach of
current computers. Furthermore, this estimate is an upper limit, as the current
\WMAP\ module has not yet been heavily optimized, and in particular a new
BiCG-STAB preconditioner may result in a significant speedup. We have also
found that the amount of recoding effort required to generalize the existing
\commander\ machinery to a new dataset is fully manageable; in this specific
case, it corresponded to $\mathcal O(1)$ postdoc~years, starting from no working
knowledge of either the \WMAP\ or \commander\ pipelines.

The second goal was to assess the quality of the maps; is the current code
ready for production work? In this case, we conclude with a tentative positive
answer.  The remaining work either consists of optimization (e.g., better
preconditioning in mapmaking, sidelobe interpolation speedups) or improving 
suboptimal data treatment (e.g., proper baseline fitting, expanded PSD model).
Indeed, our preliminary work comparing the \WMAP\ \Q1 and \Q2 sky maps suggests
that the software is quickly reaching a level of maturity at which a joint
\Planck\ LFI/\WMAP\ analysis may be performed.

Considering the new results presented in this paper, sidelobe contamination in
general has taken on a new importance with respect to the publicly available
\WMAP\ large-scale polarization data.  Although the polarized sidelobe pattern is
morphologically very similar to the imbalance modes, its contribution is
not explicitly being marginalized over in the low-resolution covariance
matrices used for low-$\ell$ CMB likelihood estimation \citep{hinshaw2012}, and
can therefore directly bias estimates of, for instance, the reionization
optical depth derived from \WMAP\ polarization data, at a level of $\mathcal
O(0.5\,\mathrm{\mu K})$.  The potential sidelobe contamination is an important
issue for future joint analyses of \WMAP\ and \Planck\ data, which currently
use pre-pixelized sky maps as in the \BP\ analysis; a nonnegligible fraction of
the \Planck--\WMAP\ residuals reported by \citet{planck2016-l04} and
\citet{bp01} may be due to this specific issue.

We argue that the main takeaway from this work is another illustration of the
importance of joint multi-experiment analysis. As amply illustrated through
both the \WMAP\ and \Planck\ analysis efforts, any given experiment has blind
spots to which they are not sensitive. These blind spots lead to unconstrained
modes in the frequency maps, which in turn may bias both astrophysical and
cosmological conclusions. We argue that the optimal solution to this problem is
not primarily more clever algorithms (even though such certainly can help), but
rather adding more data. Whenever a given degeneracy limits the data analysis,
whether it is foreground uncertainties caused by a limited frequency range, or
unconstrained map modes caused by the scanning strategy, the best solution is
to bring in more data to break the degeneracy.  This is the goal of the
\cosmoglobe\ project; to analyze the world's best data jointly. This paper is
an important step in that direction, aiming to combine the world's two best CMB
satellite datasets within one joint framework.

\begin{acknowledgements}
  We thank Prof.~Charles Bennett, Dr.~Janet Weiland, Prof.~Lyman Page, and Dr.~Zhilei Xu for useful comments and suggestions that improved this work. We thank the anonymous referee whose suggestions improved this paper.
  We thank Prof.\ Pedro Ferreira and Dr.\ Charles Lawrence for useful suggestions, comments and 
  discussions. We also thank the entire \Planck\ and \WMAP\ teams for
  invaluable support and discussions, and for their dedicated efforts
  through several decades without which this work would not be
  possible. The current work has received funding from the European
  Union’s Horizon 2020 research and innovation programme under grant
  agreement numbers 776282 (COMPET-4; \BP), 772253 (ERC;
  \textsc{bits2cosmology}), and 819478 (ERC; \textsc{Cosmoglobe}). In
  addition, the collaboration acknowledges support from ESA; ASI and
  INAF (Italy); NASA and DoE (USA); Tekes, Academy of Finland (grant
   no.\ 295113), CSC, and Magnus Ehrnrooth foundation (Finland); RCN
  (Norway; grant nos.\ 263011, 274990); and PRACE (EU).
  We acknowledge the use of the Legacy Archive for Microwave Background Data
  Analysis (LAMBDA), part of the High Energy Astrophysics Science Archive Center
  (HEASARC). HEASARC/LAMBDA is a service of the Astrophysics Science Division at
  the NASA Goddard Space Flight Center.  
\end{acknowledgements}

\bibliographystyle{aa}

\bibliography{Planck_bib,BP_bibliography}

\end{document}